\documentclass{aa}
\usepackage{graphicx}
\begin{document}

\title{Na-O Anticorrelation And HB. III. The abundances of \object{NGC 6441} from FLAMES-UVES spectra
\thanks{Based on data collected at the European Southern Observatory with
the VLT-UT2, Paranal, Chile (ESO 073.D-0211)}}

\author{R.G.~Gratton\inst{1},
	S.~Lucatello\inst{1},
        A.~Bragaglia\inst{2},
        E.~Carretta\inst{2},
        Y.~Momany\inst{1,3},
	E.~Pancino\inst{2},
	E.~Valenti\inst{2,4}}

\offprints{R.G.~Gratton}

\institute{INAF-Osservatorio Astronomico di Padova, Vicolo dell'Osservatorio 5, 35122 Padova, ITALY
\and
           INAF-Osservatorio Astronomico di Bologna, Via Ranzani 1, 40127 Bologna, ITALY
\and
           Dipartimento di Astronoma, Universit\`a di Padova, Vicolo dell'Osservatorio 5, 35122 Padova, ITALY
\and
           Dipartimento di Astronoma, Universit\`a di Bologna, Via Ranzani 1, 40127 Bologna, ITALY
}

\date{Received: ; accepted: }

\abstract
{}
{The aim of the present work is to determine accurate metallicities for a
group of red giant branch stars in the field of the bulge Globular Cluster \object{NGC 6441}.
This is the third paper of a series resulting from a  large project aimed at determining the extent of 
the Na-O anticorrelation among Globular Cluster stars and exploring its relationship with HB morphology.}
{We present an LTE abundance analysis of these objects, based on data gathered with the FLAMES fiber facility and the UVES 
spectrograph at VLT2.}
{Five of the thirteen stars observed are members of the cluster. The average Fe 
abundance for these five stars is [Fe/H]=$-0.39\pm 0.04\pm 0.05$~dex, where the first error bar includes 
the uncertainties related to star-to-star random errors, and the second one the systematic effects related 
to the various assumptions made in the analysis.The overall abundance pattern is quite typical of Globular
 Clusters, with an excess of the $\alpha-$elements and of Eu. There is evidence that the stars of \object{NGC 6441} 
are enriched in Na and Al, while they have been depleted of O and Mg, due to H-burning at high temperatures, 
in analogy with extensive observations for other Globular Clusters: in particular, one star is clearly Na and
 Al-rich and O and Mg-poor. We obtained also quite high V abundances, but it is possible that this is an artifact 
of the analysis, since similar large V abundances are derived also for the field stars. These last are all more
 metal-rich than \object{NGC 6441} and probably belong to the bulge population.
}
{}
\keywords{ Stars: abundances -
           Stars: atmospheres -
           Stars: Population II -
           Galaxy: Globular Clusters: general -
           Galaxy: Globular Clusters: individual: \object{NGC 6441} }

\authorrunning{Gratton R.G. et al.}
\titlerunning{Metallicity of \object{NGC 6441}}

\maketitle

%

\section{Introduction}

\object{NGC 6441} is a very luminous ($M_V=-9.18$: Harris 1996) and massive
Globular Cluster located in the inner regions of our Galaxy: according
to Harris (1996) it is located about 2.1 kpc from the Galactic
center on the other side of the Galaxy with respect to the
Sun. \object{NGC 6441} has raised a considerable interest because, in spite of
being a metal-rich cluster ([Fe/H]=$-0.53\pm 0.11$, Armandroff \& Zinn
1988), it has a very extended horizontal branch, with several stars on
the blue side of the instability strip (Rich et al. 1997). Layden et
al. (1999) estimated that about 17\% of the horizontal branch stars
are found in a feature bluer and brighter than the red clump, a
peculiarity shared with the other massive, metal-rich bulge cluster
\object{NGC 6388} (Rich et al. 1997). \object{NGC 6441} has also a quite rich and peculiar
population of RR Lyrae (Layden et al. 1999; Pritzl et al. 2001),
characterized by long periods in spite of the high metallicity of the
Cluster. The color-magnitude diagram of \object{NGC 6441} presents other
peculiarities: the red giant sequence is broad, and the clump is
clearly tilted. These features may at least in part be justified by
differential reddening (see e.g. Layden et al. 1999). However, given
that \object{NGC 6441} is bright and massive, some star-to-star spread in the
metal abundances cannot be excluded. Such a possibility was raised in
order to explain the odd properties of the horizontal branch and of
the RR Lyrae population (Piotto et al. 1997; Sweigart 2001; Pritzl et
al. 2001); however, very recently Clementini et al. (2005) have shown
that most of the RR Lyrae of \object{NGC 6441} are metal-rich, likely in
accordance to the other cluster stars. The peculiarities of \object{NGC 6441}
must then still find a comprehensive explanation.

All the previous discussion seems to imply that \object{NGC 6441} is
further evidence of the second parameter effect in Globular
Clusters; here, like in \object{NGC 2808}, the second parameter effect is
manifest within the same cluster. A peculiarity of \object{NGC 6441} (and of
\object{NGC 6388}) is that the blue horizontal branch is brighter than the red
one. This might be explained by a higher content in Helium (Sweigart
\& Catelan 1998). The existence of He-rich populations have been
recently proposed to justify various features of some Globular
Clusters (D'Antona \& Caloi 2003; Bedin et al. 2004; Piotto et
al. 2005; D'Antona et al. 2005). In particular, a He-rich population
might indicate a second generation of stars born from the ejecta of
the most massive of the intermediate-mass stars of the earlier
generation, required to explain the observations of O-Na and Mg-Al
abundance anticorrelations, ubiquitous in Globular Clusters but not
observed among field stars (see the review by Gratton et al. 2004). If
this scenario is correct, there should be a correlation between the
distribution of the O/Na abundance ratios, observed e.g. on the red
giant branch of a cluster, and the color distribution of stars along
the horizontal branch of the same cluster.

In order to explore whether such a correlation indeed exists, we have
undertaken an extensive spectroscopic survey of Globular Cluster stars
with the FLAMES multi-fiber facility at ESO (Pasquini et
al. 2002). With its high multiplexing capabilities, FLAMES can
get spectra of hundreds of stars in Globular Clusters within a few
hours of VLT observing time. FLAMES may simultaneously feed two
spectrographs: GIRAFFE, which is most suited for observation of
extensive samples of stars in restricted wavelength regions, and UVES,
which provides extensive spectral coverage for a few stars. The latter
higher resolution (R$\simeq$40,000) spectra may be used to carefully derive the chemical
composition of the cluster, and possibly discover the existence of
peculiarities due to its chemical evolution history, by comparing the
abundance pattern with that obtained for field stars of similar
characteristics.

\object{NGC 6441} was an obvious candidate to be included in our sample. In this
paper, we present an analysis of the FLAMES+UVES data we gathered for
stars in this cluster, as well as for a few stars in the same field
that turned out not to be members of the cluster. A separate paper
will be devoted to the discussion of the GIRAFFE data.

\section{Observations}

\begin{table*}
\begin{center}
\caption{Journal of observations}
\begin{tabular}{lccccc}
\hline
Fiber         &    Date    &   Time   & Exp. Time & Seeing   & Airmass \\ 
Configuration &            &   (UT)   & (sec)     & (arcsec) &         \\
\hline
\#1           & 2004-07-06 & 04:21:42 & 5300 & 1.59 & 1.040-1.174 \\
              & 2004-07-11 & 02:48:19 & 5300 & 1.35 & 1.029-1.052 \\ 
              & 2004-07-11 & 04:19:18 & 5300 & 0.91 & 1.055-1.221   \\ 
\#2           & 2004-07-17 & 05:18:20 & 5300 & 0.69 & 1.202-1.605 \\ 
              & 2004-07-26 & 03:39:54 & 5300 & 1.04 & 1.077-1.282 \\ 
\hline
\end{tabular}
\label{t:journal}
\end{center}
\end{table*}

Data used  in this paper are based  on exposures obtained in July 2004
with FLAMES  at Kueyen (=VLT2) used   in service mode. Details  of the
observations are given in Table~\ref{t:journal}.

\object{NGC 6441} is a quite far and concentrated cluster ($c=1.85$: Harris
1996) lying in a very  crowded region, close  to the direction of  the
Galactic center   ($l=353.53^{\circ}$, $b=-5.01^{\circ}$).  Unluckily,
when we  started our project  there was  no suitable  membership study
available. These facts made   identification of stars appropriate  for
abundance analysis     difficult, given   the safety   constraints  on
fiber-to-fiber separation of the  Oz-Poz FLAMES fiber positioner,  and
the need to avoid stars whose images are  blended with neighbors.  Our
targets were carefully selected from high quality photometry.

The photometric data have  been obtained within the observing  program
69.D-0582(A) on June  2002 at the   2.2~m ESO/MPI telescope at  La Silla
Observatory  (Chile),  using the  optical  camera WFI. The exceptional
capabilities of this imager provide a mosaic of  eight CCD chips, each
with a  field of view of ${\sim}8'{\times}16'$,  for  a total field of
view of   ${\sim}34'{\times}33'$.   The  cluster center   was  roughly
centered on Chip~\#7, and the  images were taken  through the $V$  and
$I_c$ broad  band filters with  typical  exposure times of  80~sec and
50~sec, respectively.  During the night the  average  seeing was quite
good (full width half maximum FWHM${\approx}0\farcs9$).

Stellar photometry was performed   using   the standard  and    tested
DAOPHOT~II and ALLFRAME   programs (Stetson  1994).  In   crowded
fields such as that surrounding NGC~6441,  these programs provide high
quality stellar photometry by point spread function (PSF) fitting (see
Momany  et al.  2002, 2004).  The single PSFs were constructed
for   each  image   after a  carefull   selection    of  isolated  and
well-distributed stars across each chip.
The instrumental PSF magnitudes  of   NGC~6441   were
normalized to 1~s  exposure and zero airmass, and corrected  for
the aperture correction.  The aperture corrections are necessary
to convert the instrumental  profile-fitting magnitudes of NGC~6441 on
the scale of  the  standard   star  measurements (based on    aperture
photometry,  see  below).   The instrumental NGC~6441  magnitudes  are
therefore:

\begin{equation}
m^{'}=m_{\rm apert.}+2.5\log(t_{exp})-K_{\lambda}X
\end{equation}

where  $m_{\rm apert.}=m_{\rm PSF}-{\rm apert.~correction}$, $X$
is the airmass of the reference image for each filter, and the adopted
mean   extinction coefficients   for  La  Silla  are:  $K_V=0.16$  and
$K_I=0.07$.
The aperture corrections   were  estimated upon  isolated  bright
stars, uniformally distributed across the chip.  For these we obtained
aperture photometry   after   subtracting eventual  nearby  companions
within 5 FWHM.  The aperture magnitudes measured in circular apertures
of  6 arcseconds in  radius   (close  to the photoelectric    aperture
employed by Landolt  1992) were then  compared to the  PSF magnitudes,
and  the difference is assumed to  be the aperture correction to apply
to the PSF magnitudes.

Separately, aperture magnitudes of  standard stars from 3 Landolt
(1992) fields  were measured in  circular apertures of 6 arcseconds in
radius.  These  aperture   magnitudes    were  normalized   to   their
corresponding airmass and    exposure times, and compared with   those
tabled in  Landolt (1992). A  least  square fitting procedure provided
the calibration relations.
The $r.m.s.$ scatter of the residuals of the fit ($0.009$ and $0.011$)
are assumed to represent our calibration  uncertainties in $V$ and $I$
respectively.
The total zero-point uncertainties, including the aperture corrections
and calibration errors,  are $0.013$, and  $0.016$ mag in  $V$ and $I$
respectively.  The PSF photometry  of  NGC~6441 (normalized and
corrected   for aperture correction)  was   then calibrated using  the
relations:

\begin{equation}
V=v^{'}-0.077(V-I)+24.199 
\end{equation}
\begin{equation}
I=i^{'}+0.097(V-I)+23.142
\end{equation}


The {\it  Guide  Star   Catalog  (GSCII)}  was   used to  search   for
astrometric standards  in the entire  WFI image field of view. Several
hundred  astrometric  {\it GSCII} reference   stars were found in each
chip, allowing us an accurate absolute  positioning of the sources. An
astrometric solution has been obtained for each of the eight WFI chips
independently,  by      using    suitable  catalog    matching     and
cross\--correlation tools developed at the Bologna Observatory. At the
end of  the   entire   procedure,  the    {\it  rms}  residuals    are
${\leq}$$0\farcs$2 both in right ascension and declination.

%
%
%

Only  uncrowded stars  were  considered as possible   targets, that is
stars not showing any  companion brighter than  $V_{\rm target}+2$~mag
within a 2.5~arcsec   radius, or brighter than  $V_{\rm target}-2$~mag
within  10~arcsec. The targets   were  selected to   lie close to  the
cluster mean locus in the color-magnitude diagram.

The fibers feeding the UVES spectrograph were centered on stars close
to the tip of the RGB ($15.8<V<16.3$; see Figure~\ref{f:6441cmd}). We
used two different fiber configurations: in the first set, we observed
six stars, with two fibers dedicated to the sky; in the second one,
seven fibers were used for the stars, and one for the sky. The spectra
cover the wavelength range 4700-6900~\AA, and have $35<S/N<85$ for the combined 
exposures.

As part of the service mode observations, the spectra were reduced by
ESO personnel using the dedicated UVES-FLAMES pipeline (uves/2.1.1
version). We found that this UVES pipeline does not accurately
subtract the background between orders in the green-yellow part of the
spectra. Luckily, the redder part of the spectra included enough lines
for our purposes. We hence relied on these pipelines reductions; the
bluer portions of the spectra were then not considered in the present
analysis.

\section{Cluster membership}

\begin{table*}
\begin{center}
\caption{Photometry and spectroscopic data for stars observed with UVES}
\begin{tabular}{lcccccccccccc}
\hline
Star & Fiber &  RA     &  Dec.  &  Dist. & V  & V-I  & V-K  & RV & S/N & $\sigma$(EW) & [Fe/H] & Notes \\
     & Conf. &(degree) &(degree)&(arcsec)&(mag)&(mag)&(mag)&(km s$^{-1})$& & (m\AA)  &        & \\
\hline
6003734 & 1 & 267.422 & $-$36.9533 & 518 & 15.992 & 2.083 & 4.574 &  ~+83.5& 39 &~8.7 & $-$0.25 &          \\ 
6004360 & 2 & 267.475 & $-$37.1471 & 414 & 16.127 & 1.979 & 4.403 &  ~+89.7& 37 &~8.1 &  ~~0.27 &          \\ 
6005308 & 1 & 267.463 & $-$37.0816 & 284 & 16.290 & 1.973 &       &$-$144.1& 36 &~8.5 & $-$0.17 &          \\ 
6005341 & 2 & 267.447 & $-$36.9682 & 429 & 16.294 & 2.095 & 4.543 &~$-$41.2& 50 &~9.4 &  ~~0.06 &          \\ 
7003717 & 2 & 267.593 & $-$36.9987 & 219 & 15.989 & 2.113 & 4.522 &$-$134.0& 50 &     &         & fast rot.\\
7004050 & 1 & 267.543 & $-$37.0793 & 106 & 16.062 & 2.168 & 4.775 &  ~+24.9& 52 &~8.0 & $-$0.45 & member   \\
7004329 & 1 & 267.581 & $-$37.1614 & 404 & 16.120 & 1.997 & 4.258 &~$-$23.1& 51 &10.0 &  ~~0.12 &          \\ 
7004434 & 2 & 267.541 & $-$37.0849 & 127 & 16.142 & 2.167 & 4.719 &  ~+13.6& 53 &10.0 & $-$0.34 & member   \\
7004453 & 1 & 267.525 & $-$37.0109 & 167 & 16.144 & 2.029 & 4.382 &~$-$49.3& 49 &~6.7 & $-$0.04 &          \\ 
7004463 & 2 & 267.506 & $-$37.0572 & 140 & 16.146 & 2.160 &       &  ~+11.6& 77 &~9.1 & $-$0.38 & member   \\
7004487 & 2 & 267.520 & $-$37.0535 & ~99 & 16.150 & 2.088 & 4.552 &  ~+32.3& 82 &~6.1 & $-$0.50 & member   \\
8002961 & 2 & 267.670 & $-$37.0725 & 342 & 15.798 & 2.185 & 4.803 &  ~+54.1& 86 &~5.8 & $-$0.28 & member?  \\ 
8003092 & 1 & 267.753 & $-$37.0622 & 573 & 15.832 & 2.203 & 4.923 &~$-$53.0& 37 &~7.8 & $-$0.25 &          \\
\hline
\end{tabular}
\label{t:uvesphot}
\end{center}
\end{table*}

Table \ref{t:uvesphot} gives details on the main parameters for the
observed stars, providing for each one of them 
the distance in arcsec from the cluster center, 
the magnitudes, mean radial velocities, signal-to-noise ratios of the 
combined spectra and the $\sigma$ of the EWs residuals, as well as the derived 
metallicity.
 Star designations are according to Valenti et al. (2006), from which
photometric data were also taken. Coordinates (at J2000 equinox) are
from our astrometry (Valenti et al. 2006); distances from the cluster center were
obtained considering the nominal position given by Harris
(1996). Radial velocities were measured from our spectra, using
typically about 100 atomic lines. Errors in these velocities should typically be
of a few hundreds of m/s. The signal-to-noise ratio S/N was estimated
from the pixel-to-pixel scatter in spectral regions free from
absorption lines at about 6200~\AA.

\begin{figure}[h]
\includegraphics[width=8.8cm]{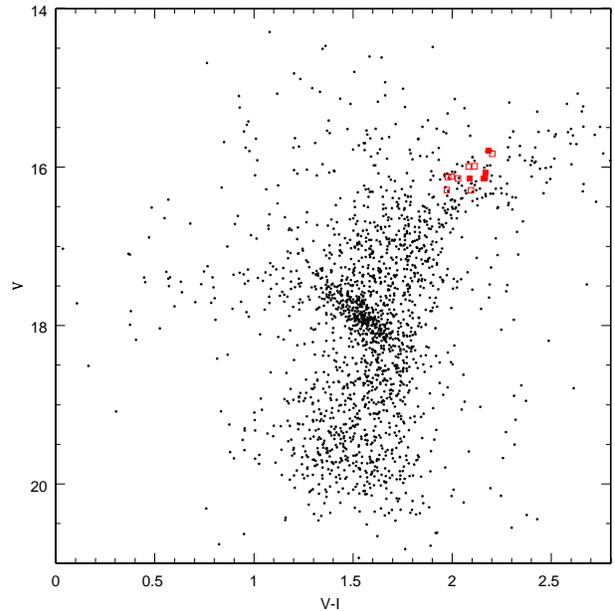}
\caption[]{$(V,V-I)$ color magnitude diagram for selected stars in the field of \object{NGC 6441} 
(from Valenti et al. 2006). Large symbols are stars observed with UVES FLAMES. Filled symbols 
mark stars member of the cluster, on the basis of radial velocities and location in 
the field close to the cluster center; open symbols are non-member stars.}
\label{f:6441cmd}
\end{figure}

Four of the thirteen observed stars (\#7004050, 7004434, 7004463, and
7004487) are very likely members of the cluster, on the basis of the
following criteria:
\begin{itemize}
\item they are projected close to the nominal cluster center, as given by Harris (1996).
These four stars are the closest in our sample to the nominal cluster center, 
and are all within 150 arcsec from this position. Note that according to Harris (1996) 
the core radius of \object{NGC 6441} is about 7 arcsec, and the tidal radius is about 467 arcsec.
\item their radial velocities (+24.9, +13.6, +11.6, and +32.3 km~s$^{-1}$\ for stars 7004050,
 7004434, 7004463, and 7004487, respectively) are close to the average radial velocity 
listed by Harris ($+16.4\pm 1.2$~km~s$^{-1}$). The average velocity given by these four 
stars is $+20.6\pm 4.9$~km~s$^{-1}$\ ($r.m.s.$ scatter of 9.8 km~s$^{-1}$), in good 
agreement with the value listed by Harris.
\item additionally, the four stars are confined in a very small range of colors, with 
$2.088<V-I<2.168$, with respect to the rather broad color distribution along the apparent 
RGB of \object{NGC 6441} (see Figure~\ref{f:6441cmd}). This by itself is not a strong membership 
criterion, since the interstellar reddening is expected to be quite variable over the field 
(see Layden et al. 1999 for an extensive discussion). However, the close agreement between 
the colors of the confirmed member stars might indicate that differential reddening is less of
 a problem than expected, at least for these particular stars (it should be noted that all 
these stars are in the south-east quadrant of the cluster, quite close to each other).
\item a posteriori, our analysis indicates that these stars have quite similar chemical 
composition. Again, this is not by itself a strong membership criterion, since there
 might be some spread in metallicity amongst the stars of \object{NGC 6441} and because 
their metallicity is not unusual for bulge or disk field stars.   
\end{itemize}

Among the remaining stars, \#8002961 is also a possible cluster
member, essentially on the basis of the not too large difference
between its radial velocity (+54.1 km~s$^{-1}$) and that of the
cluster\footnote{While the velocity difference might appear large 
 we note that \object{NGC 6441} is well known to have an extremely large central velocity 
dispersion (around 17\,km s$^{-1}$ see e.g. Dubath et al. 1997).}, 
and its location in the color magnitude diagram, since it
lies close to the position occupied by the member stars. This star is
projected at a quite large distance from the cluster center (341
arcsec), however still well within the tidal radius of \object{NGC 6441}. A
posteriori, also the chemical composition supports cluster membership.
It is noteworthy that the inclusion or exclusion of this particular 
stars from the group ob objects considered cluster members does not 
affect our conclusions.

All remaining stars are unlikely to be cluster members, because of
their discrepant radial velocities, large projected distances from the
cluster center, much bluer colors, and often quite discrepant
chemical composition. The rather large fraction of non-members did not
come as a surprise, given the difficulties in identifying good
candidates in this cluster and the strong contamination by field bulge
stars. We notice that only observations by means of a multi-object
facility like FLAMES could ensure the success of the program.

There are various circumstantial arguments that favor the membership
to the bulge of all remaining stars. Three of them
(\#6005341, 7004329, and 7004453) have small negative radial
velocities ($-$41.2, $-$23.1, and $-$49.3 km~s$^{-1}$, respectively)
and nearly solar Fe abundance. However, given the location of \object{NGC 6441}
at $b=-5.01$, they should be within 2-3 kpc from the Sun to be members
of the thin disk (i.e. at no more than 200-300 pc from the galactic
plane: see e.g. Bilir et al. 2005). Were this the correct distance
of the stars, they would  be sub-giants, with a surface gravity
of about $\log g\sim 3$\footnote{This combination of effective
temperatures and surface gravity would also be unlikely by
evolutionary considerations, since sub-giants are much warmer than
$\sim 4000$~K for any reasonable chemical composition. Only pre-main
sequence stars could occupy a similar location on the color-magnitude
diagram}. This is clearly excluded by their spectra, that indicate a
much lower surface gravity of $\log g<2$. These stars are then at more
than 6 kpc from the Sun, and at more than 500 pc from the galactic
plane. The same surface gravity argument can be applied to the
remaining stars. We also note that stars \#6003734 and \#8003092 have
a chemical composition similar to that of \object{NGC 6441}, but have very
different radial velocities (+83.5 and $-53.0$~km~s$^{-1}$) and are
projected far from the cluster center (518 and 573 arcsec
respectively, more than the tidal radius). Also star \#6003734 is
distinctly bluer than the cluster members considered above, while star
\#8003092 has magnitude and color very similar to those of star
\#8002961, that we considered a possible member. Star \#6005308 has a
chemical composition similar to that of \object{NGC 6441}: however its
radial velocity is very different ($-$144.1 km~s$^{-1}$), clearly
incompatible with membership. Finally, star \#6004360 has a large
positive velocity of 89.7 km~s$^{-1}$ and a
large metal content. All these stars are distinctly
bluer than the cluster members, although more
metal-rich. This might be due to one of these three causes: smaller
interstellar reddening (however there is no clear indication of this
from our analysis); shorter distances (putting the stars close to the
galactic center), or younger ages. As we will see, a shorter distance
by about 20\% is also suggested by the equilibrium of ionization. We
conclude that all these stars most likely belong to the bulge population.

Finally, one of the target stars (\#7003717) turned out to have very
broad lines, likely widened by rotation. We measured a large negative
radial velocity ($-$134.0 km~s$^{-1}$), with clear indications of
quite large variations over the small temporal range covered by our
observations. Due to difficulties in the analysis, we did not further
consider this star.

\begin{table*}
\caption{Equivalent Widths from UVES spectra for the cluster members (in electronic form)}
\scriptsize
\begin{tabular}{lccccccccccccc}
\hline
El.  & Wavel.  & E.P. & log gf &  4050 &      &  4434 &      &  4463 &      &  4487 &      &  2961 &      \\
     &         &      &        & EW    & log A& EW    & log A& EW    & log A& EW    & log A&EW     & log A\\
     & (\AA)   & (eV) &        & (m\AA)&      & (m\AA)&      & (m\AA)&      & (m\AA)&      &(m\AA) &      \\
\hline
 O{\sc i}& 6300.31 & 0.00 & $-$9.75 & ~56.2 & 8.23 & ~83.9 & 8.61 &       &      & ~63.4 & 8.45 & ~68.7 & 8.39 \\
 O{\sc i}& 6363.79 & 0.02 &$-$10.25 & ~29.7 & 8.44 & ~50.9 & 8.76 & ~15.8 & 8.14 & ~38.5 & 8.65 & ~33.3 & 8.44 \\
Na{\sc i}& 6154.23 & 2.10 & $-$1.57 & ~99.4 & 6.36 & 105.4 & 6.14 & 140.8 & 6.62 & 100.4 & 6.20 & 110.5 & 6.34 \\
Na{\sc i}& 6160.75 & 2.10 & $-$1.26 & 111.1 & 6.25 & 137.5 & 6.24 & 154.4 & 6.49 & 123.3 & 6.19 & 136.2 & 6.40 \\
Mg{\sc i}& 6318.71 & 5.11 & $-$1.94 & ~59.1 & 7.30 & ~73.9 & 7.38 & ~61.7 & 7.20 & ~75.1 & 7.41 & ~77.8 & 7.51 \\
Mg{\sc i}& 6319.24 & 5.11 & $-$2.16 & ~48.2 & 7.31 & ~64.6 & 7.46 & ~48.5 & 7.22 & ~53.0 & 7.29 & ~70.8 & 7.61 \\
Al{\sc i} & 6696.03 & 3.14 & $-$1.32 & ~99.9 & 6.30 & 128.3 & 6.40 & 164.2 & 6.95 & 106.6 & 6.22 & 127.4 & 6.54 \\
Al{\sc i} & 6698.67 & 3.14 & $-$1.62 & ~77.7 & 6.18 & ~92.4 & 6.18 & 119.9 & 6.62 & ~71.1 & 6.00 & ~89.4 & 6.21 \\
Si{\sc i} & 5948.55 & 5.08 & $-$1.23 & ~67.7 & 7.41 & ~85.4 & 7.47 & 103.0 & 7.75 & ~84.4 & 7.44 & ~84.4 & 7.57 \\
Si{\sc i} & 6125.03 & 5.62 & $-$1.57 &       &      &       &      &       &      & ~27.9 & 7.35 &       &      \\
Si{\sc i} & 6145.02 & 5.62 & $-$1.49 &       &      & ~27.8 & 7.33 & ~34.4 & 7.48 & ~36.2 & 7.47 & ~45.4 & 7.78 \\
Ca{\sc i} & 5857.46 & 2.93 &  ~~0.24 & 164.0 & 5.75 & 178.9 & 5.55 & 207.6 & 5.88 & 169.2 & 5.63 & 175.7 & 5.70 \\
Ca{\sc i} & 5867.57 & 2.93 & $-$1.49 & ~56.5 & 5.59 & ~77.5 & 5.74 & ~92.8 & 6.01 & ~69.9 & 5.76 & ~77.4 & 5.82 \\
Ca{\sc i} & 6161.30 & 2.52 & $-$1.27 & 109.0 & 5.88 & 141.1 & 5.89 &       &      & 135.4 & 6.03 & 144.7 & 6.16 \\
Ca{\sc i} & 6166.44 & 2.52 & $-$1.14 & 103.8 & 5.64 &       &      &       &      & 136.5 & 5.92 & 136.0 & 5.89 \\
Ca{\sc i} & 6169.04 & 2.52 & $-$0.80 & 145.2 & 6.02 & 158.1 & 5.69 & 169.9 & 5.90 & 160.5 & 5.95 & 156.9 & 5.89 \\
Ca{\sc i} & 6169.56 & 2.52 & $-$0.48 & 136.7 & 5.58 & 176.2 & 5.62 & 187.9 & 5.80 & 146.2 & 5.42 & 169.4 & 5.74 \\
Ca{\sc i} & 6439.08 & 2.52 &  ~~0.39 & 192.1 & 5.48 & 252.0 & 5.63 & 241.6 & 5.58 & 202.6 & 5.36 & 239.7 & 5.68 \\
Ca{\sc i} & 6455.60 & 2.52 & $-$1.29 & ~93.5 & 5.56 & 145.4 & 5.98 & 137.2 & 5.89 & 114.3 & 5.68 & 125.6 & 5.85 \\
Ca{\sc i} & 6462.57 & 2.52 &  ~~0.26 & 246.0 & 5.99 & 309.9 & 6.12 & 296.6 & 6.08 & 261.3 & 6.00 & 280.9 & 6.07 \\
Ca{\sc i} & 6471.67 & 2.52 & $-$0.69 & 140.9 & 5.93 & 194.2 & 6.11 & 162.7 & 5.72 & 158.4 & 5.85 & 173.9 & 6.08 \\
Ca{\sc i} & 6493.79 & 2.52 & $-$0.11 & 166.6 & 5.72 & 209.8 & 5.73 & 209.7 & 5.77 & 180.5 & 5.59 & 198.8 & 5.82 \\
Ca{\sc i} & 6499.65 & 2.52 & $-$0.82 & 113.4 & 5.56 & 176.3 & 5.99 & 163.7 & 5.86 & 138.5 & 5.64 & 167.6 & 6.13 \\
Sc{\sc ii} & 6245.62 & 1.51 & $-$1.05 & ~71.6 & 2.72 & ~92.7 & 2.72 & 104.7 & 2.95 & ~93.4 & 2.83 & ~88.9 & 2.75 \\
Sc{\sc ii} & 6279.74 & 1.50 & $-$1.16 & ~85.8 & 3.18 & ~97.6 & 2.93 & 115.5 & 3.26 & ~81.3 & 2.75 & ~78.1 & 2.69 \\
Sc{\sc ii} & 6604.60 & 1.36 & $-$1.15 & ~91.6 & 3.03 & 115.3 & 2.94 & 108.5 & 2.86 & ~92.1 & 2.68 & 103.0 & 2.88 \\
Ti{\sc i} & 5866.46 & 1.07 & $-$0.84 & 150.8 & 4.78 & 210.1 & 4.94 & 221.4 & 5.14 & 163.7 & 4.54 & 196.5 & 5.01 \\
Ti{\sc i} & 5922.12 & 1.05 & $-$1.47 & 128.8 & 4.88 & 170.0 & 4.81 & 185.9 & 5.16 & 142.0 & 4.66 & 165.9 & 5.02 \\
Ti{\sc i} & 5978.55 & 1.87 & $-$0.50 & 111.7 & 4.77 & 137.1 & 4.54 & 148.6 & 4.81 & 122.7 & 4.58 & 133.7 & 4.74 \\
Ti{\sc i} & 6091.18 & 2.27 & $-$0.42 &       &      & 103.3 & 4.49 & 114.8 & 4.74 & 103.1 & 4.72 & 109.9 & 4.78 \\
Ti{\sc i} & 6126.22 & 1.07 & $-$1.42 & 128.6 & 4.81 & 181.9 & 4.92 & 169.9 & 4.78 & 149.5 & 4.71 & 149.2 & 4.62 \\
Ti{\sc i} & 6258.11 & 1.44 & $-$0.36 & 176.6 & 4.99 & 220.7 & 4.95 & 223.2 & 5.04 & 188.3 & 4.86 & 187.4 & 4.80 \\
Ti{\sc i} & 6554.24 & 1.44 & $-$1.22 &                                                           & 151.1 & 4.91 \\
Ti{\sc ii} & 6606.98 & 2.06 & $-$2.90 & ~41.6 & 5.14 & ~45.4 & 5.07 & ~32.8 & 4.84 & ~29.8 & 4.77 & ~51.4 & 5.19 \\
V{\sc i} & 6002.31 & 1.22 & $-$1.77 & ~59.2 & 3.76 & ~63.8 & 3.64 & ~79.3 & 3.87 & ~70.9 & 3.92 & ~58.2 & 3.56 \\
V{\sc i} & 6039.73 & 1.06 & $-$0.65 &       &      & 144.5 & 3.43 & 159.8 & 3.76 &       &      & 138.3 & 3.60 \\
V{\sc i} & 6081.45 & 1.05 & $-$0.58 &       &      & 168.9 & 3.92 & 174.6 & 4.10 & 146.8 & 3.85 & 158.3 & 3.99 \\
V{\sc i} & 6090.22 & 1.08 & $-$0.06 & 133.0 & 3.62 & 193.8 & 3.83 & 183.2 & 3.76 & 158.5 & 3.59 & 172.8 & 3.79 \\
V{\sc i} & 6111.65 & 1.04 & $-$0.71 &       &      & 180.7 & 3.97 & 178.9 & 4.02 & 163.0 & 4.04 &       &      \\
V{\sc i} & 6119.53 & 1.06 & $-$0.32 & 130.3 & 3.77 & 164.9 & 3.57 & 174.2 & 3.80 & 146.6 & 3.58 & 166.2 & 3.87 \\
V{\sc i} & 6135.38 & 1.05 & $-$0.75 & 118.6 & 3.80 & 149.4 & 3.54 & 162.2 & 3.81 & 140.0 & 3.71 & 147.3 & 3.76 \\
V{\sc i} & 6199.19 & 0.29 & $-$1.28 & 175.0 & 4.04 & 217.7 & 4.01 & 224.9 & 4.19 & 187.9 & 3.90 &       &      \\
V{\sc i} & 6216.36 & 0.28 & $-$0.81 &       &      & 215.6 & 3.46 & 224.9 & 3.68 & 197.6 & 3.55 & 202.6 & 3.54 \\
V{\sc i} & 6251.82 & 0.29 & $-$1.34 &       &      & 188.5 & 3.63 & 210.5 & 4.09 & 180.9 & 3.89 &       &      \\
V{\sc i} & 6256.90 & 0.28 & $-$2.01 & 111.1 & 3.65 & 167.3 & 3.88 & 176.6 & 4.10 &       &      & 143.4 & 3.74 \\
V{\sc i} & 6274.65 & 0.27 & $-$1.67 & 121.4 & 3.58 & 168.6 & 3.54 & 174.5 & 3.72 &       &      & 158.1 & 3.65 \\
V{\sc i} & 6285.17 & 0.28 & $-$1.51 & 135.2 & 3.76 & 188.6 & 3.71 & 194.5 & 3.89 & 167.6 & 3.76 & 163.8 & 3.62 \\
V{\sc i} & 6292.83 & 0.29 & $-$1.47 & 152.5 & 4.03 & 194.6 & 3.79 & 205.5 & 4.03 & 174.3 & 3.86 & 180.9 & 3.93 \\
V{\sc i} & 6531.43 & 1.22 & $-$0.84 &       &      & 152.5 & 3.83 &       &      & 128.4 & 3.77 &       &      \\
\hline
\end{tabular}
\normalsize
\label{t:ewidths}
\end{table*}

\addtocounter{table}{-1}

\begin{table*}
\caption{Continued}
\scriptsize
\begin{tabular}{lccccccccccccc}
\hline
El.  & Wavel.  & E.P. & log gf &  4050 &      &  4434 &      &  4463 &      &  4487 &      &  2961 &      \\
     &         &      &        & EW    & log A& EW    & log A& EW    & log A& EW    & log A&EW     & log A\\
     & (\AA)   & (eV) &        & (m\AA)&      & (m\AA)&      & (m\AA)&      & (m\AA)&      &(m\AA) &      \\
\hline
Cr{\sc i} & 6330.10 & 0.94 & $-$2.87 & 139.6 & 5.52 & 189.0 & 5.51 & 173.1 & 5.34 & 144.6 & 5.11 & 157.5 & 5.28 \\
Mn{\sc i} & 6013.50 & 3.07 & $-$0.25 & 139.8 & 5.18 & 169.3 & 5.05 & 187.1 & 5.30 & 158.6 & 5.07 & 172.2 & 5.29 \\
Mn{\sc i} & 6016.65 & 3.07 & $-$0.09 & 125.3 & 4.88 & 156.5 & 4.78 & 162.3 & 4.92 & 145.9 & 4.79 & 162.2 & 5.09 \\
Mn{\sc i} & 6021.80 & 3.08 &  ~~0.03 & 122.2 & 4.83 & 165.6 & 4.91 & 174.9 & 5.07 & 141.9 & 4.73 & 163.9 & 5.11 \\
Fe{\sc i} & 5855.09 & 4.61 & $-$1.48 & ~37.3 & 7.09 &       &      & ~58.7 & 7.34 & ~44.0 & 7.08 & ~45.2 & 7.14 \\
Fe{\sc i} & 5856.10 & 4.29 & $-$1.57 & ~46.3 & 6.98 & ~72.7 & 7.23 & ~87.5 & 7.52 & ~66.3 & 7.17 & ~77.0 & 7.46 \\
Fe{\sc i} & 5858.78 & 4.22 & $-$2.19 & ~23.3 & 6.89 &       &      & ~37.9 & 7.13 & ~38.0 & 7.15 & ~36.2 & 7.12 \\
Fe{\sc i} & 5859.60 & 4.55 & $-$0.70 & ~68.6 & 7.02 & ~88.0 & 6.98 & 111.3 & 7.40 & ~90.8 & 7.10 & ~95.1 & 7.30 \\
Fe{\sc i} & 5862.37 & 4.55 & $-$0.50 & ~79.5 & 7.07 & ~99.8 & 6.99 & 122.2 & 7.39 & ~97.3 & 7.03 & 104.2 & 7.28 \\
Fe{\sc i} & 5905.68 & 4.65 & $-$0.76 & ~55.1 & 6.89 & ~83.3 & 7.10 & ~89.4 & 7.23 & ~72.2 & 6.95 & ~75.4 & 7.10 \\
Fe{\sc i} & 5927.80 & 4.65 & $-$1.06 & ~40.4 & 6.80 & ~72.7 & 7.21 & ~72.5 & 7.22 & ~62.4 & 7.06 & ~64.6 & 7.17 \\
Fe{\sc i} & 5929.68 & 4.55 & $-$1.24 & ~49.8 & 7.10 & ~65.0 & 7.12 & ~71.8 & 7.25 & ~59.7 & 7.06 & ~55.8 & 7.03 \\
Fe{\sc i} & 5930.19 & 4.65 & $-$0.29 & ~74.9 & 6.92 & 100.5 & 6.94 & 116.3 & 7.26 & ~86.4 & 6.76 & 110.0 & 7.34 \\
Fe{\sc i} & 5934.67 & 3.93 & $-$1.15 & ~92.0 & 7.17 & 107.3 & 6.90 & 136.6 & 7.44 & 110.0 & 7.08 & 117.4 & 7.32 \\
Fe{\sc i} & 5956.71 & 0.86 & $-$4.60 & 130.0 & 7.09 & 187.6 & 7.24 & 186.5 & 7.29 & 160.3 & 7.06 & 176.5 & 7.36 \\
Fe{\sc i} & 5976.79 & 3.94 & $-$1.33 &       &      & ~90.3 & 6.79 & 101.5 & 7.02 & ~97.4 & 7.03 & 101.0 & 7.20 \\
Fe{\sc i} & 5984.83 & 4.73 & $-$0.39 &       &      & 125.9 & 7.52 &       &      &       &      &       &      \\
Fe{\sc i} & 6003.02 & 3.88 & $-$1.08 & ~90.5 & 7.00 & 120.7 & 7.00 & 124.3 & 7.10 & 110.4 & 6.95 & 121.3 & 7.26 \\
Fe{\sc i} & 6007.97 & 4.65 & $-$0.82 & ~56.3 & 6.96 & ~74.4 & 6.99 & ~82.9 & 7.16 & ~81.8 & 7.18 & ~77.4 & 7.18 \\
Fe{\sc i} & 6008.57 & 3.88 & $-$0.96 & ~96.7 & 7.03 & 119.0 & 6.86 & 135.8 & 7.18 & 108.9 & 6.80 & 129.6 & 7.29 \\
Fe{\sc i} & 6027.06 & 4.08 & $-$1.23 & ~72.9 & 6.98 & 108.0 & 7.21 & 111.0 & 7.29 & 100.3 & 7.18 & ~93.9 & 7.14 \\
Fe{\sc i} & 6056.01 & 4.73 & $-$0.42 & ~65.1 & 6.88 & ~91.9 & 7.03 & ~86.6 & 6.95 & ~90.7 & 7.08 & 104.1 & 7.47 \\
Fe{\sc i} & 6079.02 & 4.65 & $-$0.95 &       &      &       &      &       &      &       &      & ~68.8 & 7.15 \\
Fe{\sc i} & 6082.72 & 2.22 & $-$3.57 & ~86.3 & 7.02 & 146.0 & 7.57 & 141.8 & 7.53 &       &      & 131.3 & 7.57 \\
Fe{\sc i} & 6089.57 & 4.58 & $-$1.28 &       &      &       &      & ~85.3 & 7.56 &       &      &       &      \\
Fe{\sc i} & 6093.65 & 4.61 & $-$1.32 & ~42.6 & 7.07 &       &      & ~60.3 & 7.20 & ~58.1 & 7.19 & ~59.1 & 7.26 \\
Fe{\sc i} & 6094.38 & 4.65 & $-$1.56 & ~34.3 & 7.15 &       &      & ~33.9 & 7.00 & ~44.1 & 7.22 &       &      \\
Fe{\sc i} & 6096.67 & 3.98 & $-$1.77 & ~58.8 & 7.06 & ~79.1 & 7.09 & ~72.3 & 6.99 & ~75.3 & 7.09 & ~75.2 & 7.14 \\
Fe{\sc i} & 6098.25 & 4.56 & $-$1.81 & ~38.5 & 7.38 &       &      &       &      &       &      &       &      \\
Fe{\sc i} & 6137.00 & 2.20 & $-$2.95 & 125.6 & 7.28 & 183.9 & 7.47 & 171.7 & 7.34 & 158.9 & 7.32 & 163.8 & 7.46 \\
Fe{\sc i} & 6151.62 & 2.18 & $-$3.30 & ~99.1 & 6.99 & 156.6 & 7.38 & 144.2 & 7.20 & 127.8 & 7.06 & 138.1 & 7.32 \\
Fe{\sc i} & 6173.34 & 2.22 & $-$2.88 & 112.4 & 6.98 & 159.7 & 7.07 & 156.2 & 7.06 & 144.2 & 7.00 & 153.9 & 7.26 \\
Fe{\sc i} & 6188.00 & 3.94 & $-$1.63 &       &      & ~85.4 & 7.00 & ~84.6 & 7.00 & ~77.9 & 6.94 & ~78.8 & 7.01 \\
Fe{\sc i} & 6200.32 & 2.61 & $-$2.44 & 106.0 & 6.94 &       &      & 146.4 & 7.02 & 132.1 & 6.90 & 136.7 & 7.07 \\
Fe{\sc i} & 6213.44 & 2.22 & $-$2.54 & 149.4 & 7.26 & 169.1 & 6.88 & 173.9 & 7.00 & 156.5 & 6.89 & 165.4 & 7.09 \\
Fe{\sc i} & 6219.29 & 2.20 & $-$2.43 & 134.5 & 6.92 & 188.3 & 7.00 & 188.9 & 7.07 & 173.4 & 7.02 & 177.4 & 7.13 \\
Fe{\sc i} & 6226.74 & 3.88 & $-$2.08 & ~46.1 & 6.89 & ~80.0 & 7.27 & ~57.6 & 6.90 & ~70.2 & 7.16 & ~61.3 & 7.02 \\
Fe{\sc i} & 6232.65 & 3.65 & $-$1.22 & ~97.7 & 6.96 & 133.3 & 7.00 & 136.3 & 7.09 & 130.3 & 7.09 & 121.0 & 7.03 \\
Fe{\sc i} & 6240.65 & 2.22 & $-$3.23 & ~95.8 & 6.89 & 131.9 & 6.93 & 144.8 & 7.20 & 121.3 & 6.93 & 130.8 & 7.17 \\
Fe{\sc i} & 6265.14 & 2.18 & $-$2.55 & 147.4 & 7.19 & 208.5 & 7.34 & 209.2 & 7.39 & 181.6 & 7.22 & 187.4 & 7.34 \\
Fe{\sc i} & 6270.23 & 2.86 & $-$2.46 &       &      & 121.3 & 6.92 & 121.5 & 6.96 &       &      & 114.7 & 7.03 \\
Fe{\sc i} & 6297.80 & 2.22 & $-$2.74 & 146.7 & 7.42 & 191.9 & 7.39 & 185.0 & 7.36 & 159.4 & 7.14 & 180.7 & 7.50 \\
Fe{\sc i} & 6301.51 & 3.65 & $-$0.72 & 116.7 & 6.80 & 161.0 & 6.89 & 158.2 & 6.88 &       &      & 164.1 & 7.15 \\
Fe{\sc i} & 6311.50 & 2.83 & $-$3.16 & ~76.0 & 7.19 & 115.6 & 7.44 & 107.7 & 7.33 &       &      & 101.3 & 7.38 \\
Fe{\sc i} & 6315.81 & 4.08 & $-$1.68 & ~46.0 & 6.76 & ~78.4 & 7.12 & ~72.2 & 7.02 & ~61.0 & 6.86 & ~72.2 & 7.12 \\
Fe{\sc i} & 6322.69 & 2.59 & $-$2.43 & 119.2 & 7.16 & 154.8 & 7.04 & 144.8 & 6.90 & 137.3 & 6.92 & 145.7 & 7.16 \\
Fe{\sc i} & 6330.85 & 4.73 & $-$1.22 & ~52.0 & 7.37 &       &      & ~59.0 & 7.24 & ~49.8 & 7.08 & ~70.7 & 7.56 \\
Fe{\sc i} & 6335.34 & 2.20 & $-$2.27 &       &      & 220.2 & 7.17 &       &      &       &      & 193.5 & 7.10 \\
Fe{\sc i} & 6380.75 & 4.19 & $-$1.37 &       &      & ~96.3 & 7.27 & ~92.7 & 7.23 & ~80.9 & 7.07 & ~97.7 & 7.49 \\
Fe{\sc i} & 6392.54 & 2.28 & $-$3.97 & ~78.3 & 7.25 & 107.6 & 7.30 & ~83.8 & 6.92 &       &      & ~90.3 & 7.14 \\
Fe{\sc i} & 6400.32 & 3.60 & $-$0.23 &       &      & 263.3 & 7.28 &       &      &       &      & 225.1 & 7.17 \\
Fe{\sc i} & 6411.66 & 3.65 & $-$0.60 & 129.7 & 6.87 & 172.7 & 6.92 &       &      &       &      & 164.8 & 7.04 \\
Fe{\sc i} & 6421.36 & 2.28 & $-$2.03 &       &      & 235.4 & 7.16 &       &      &       &      &       &      \\
Fe{\sc i} & 6481.88 & 2.28 & $-$2.98 & 119.3 & 7.26 & 168.3 & 7.34 & 152.2 & 7.13 & 140.4 & 7.08 & 158.7 & 7.47 \\
\hline
\end{tabular}
\normalsize
\end{table*}

\addtocounter{table}{-1}

\begin{table*}
\caption{Continued}
\scriptsize
\begin{tabular}{lccccccccccccc}
\hline
El.  & Wavel.  & E.P. & log gf &  4050 &      &  4434 &      &  4463 &      &  4487 &      &  2961 &      \\
     &         &      &        & EW    & log A& EW    & log A& EW    & log A& EW    & log A&EW     & log A\\
     & (\AA)   & (eV) &        & (m\AA)&      & (m\AA)&      & (m\AA)&      & (m\AA)&      &(m\AA) &      \\
\hline
Fe{\sc i} & 6498.94 & 0.96 & $-$4.70 &       &      & 216.4 & 7.73 & 185.8 & 7.37 & 162.0 & 7.22 & 184.3 & 7.59 \\
Fe{\sc i} & 6518.37 & 2.83 & $-$2.46 & 114.0 & 7.40 & 151.5 & 7.34 & 125.1 & 6.92 & 120.9 & 6.97 &       &      \\
Fe{\sc i} & 6533.94 & 4.56 & $-$1.29 &       &      & ~89.4 & 7.58 & ~58.1 & 7.05 & ~49.0 & 6.90 & ~68.6 & 7.33 \\
Fe{\sc i} & 6574.25 & 0.99 & $-$5.00 & 125.7 & 7.46 & 183.6 & 7.60 & 150.3 & 7.12 & 133.9 & 7.01 & 150.7 & 7.34 \\
Fe{\sc i} & 6581.22 & 1.49 & $-$4.68 &       &      & 157.5 & 7.63 & 145.5 & 7.48 & 119.8 & 7.19 & 134.4 & 7.48 \\
Fe{\sc i} & 6593.88 & 2.43 & $-$2.42 & 137.6 & 7.17 & 186.3 & 7.21 & 166.6 & 6.98 & 153.2 & 6.92 & 175.1 & 7.30 \\
Fe{\sc i} & 6608.04 & 2.28 & $-$3.96 & ~73.0 & 7.07 & 115.8 & 7.39 & ~99.8 & 7.16 & ~87.4 & 7.05 & 105.7 & 7.41 \\
Fe{\sc i} & 6609.12 & 2.56 & $-$2.69 &       &      & 165.3 & 7.38 & 153.5 & 7.24 & 142.3 & 7.19 & 150.5 & 7.42 \\
Fe{\sc i} & 6627.56 & 4.55 & $-$1.50 &       &      &       &      & ~45.4 & 7.02 & ~38.1 & 6.89 & ~61.9 & 7.39 \\
Fe{\sc i} & 6633.76 & 4.56 & $-$0.82 & ~74.6 & 7.28 & 101.6 & 7.33 & ~80.1 & 6.97 & ~82.3 & 7.05 & ~96.5 & 7.43 \\
Fe{\sc i} & 6703.58 & 2.76 & $-$3.01 & ~92.7 & 7.30 & 128.1 & 7.34 & 108.6 & 7.04 & ~91.0 & 6.84 & 109.6 & 7.23 \\
Fe{\sc i} & 6713.75 & 4.80 & $-$1.41 & ~41.3 & 7.37 &       &      &       &      &       &      &       &      \\
Fe{\sc i} & 6725.36 & 4.10 & $-$2.21 & ~45.4 & 7.29 &       &      & ~35.5 & 6.92 &       &      & ~49.8 & 7.22 \\
Fe{\sc i} & 6726.67 & 4.61 & $-$1.10 & ~43.9 & 6.86 & ~70.7 & 7.12 & ~60.4 & 6.96 & ~54.3 & 6.87 & ~71.0 & 7.24 \\
Fe{\sc i} & 6733.15 & 4.64 & $-$1.44 & ~42.7 & 7.21 &       &      & ~42.4 & 7.02 & ~35.1 & 6.88 & ~50.5 & 7.23 \\
Fe{\sc i} & 6750.16 & 2.42 & $-$2.62 & 124.7 & 7.09 & 181.9 & 7.28 & 161.3 & 7.03 & 146.3 & 6.94 & 157.9 & 7.20 \\
Fe{\sc i} & 6786.86 & 4.19 & $-$1.90 & ~52.7 & 7.28 & ~68.4 & 7.30 & ~49.8 & 6.99 &       &      & ~61.1 & 7.26 \\
Fe{\sc ii} & 6247.56 & 3.89 & $-$2.33 & ~24.0 & 6.91 &       &      & ~19.2 & 6.50 & ~31.7 & 6.86 & ~28.7 & 6.92 \\
Fe{\sc ii} & 6432.68 & 2.89 & $-$3.58 & ~22.1 & 6.80 & ~48.2 & 7.35 & ~25.3 & 6.74 & ~26.3 & 6.70 & ~42.6 & 7.29 \\
Fe{\sc ii} & 6456.39 & 3.90 & $-$2.10 & ~33.3 & 7.07 & ~61.5 & 7.52 & ~38.2 & 6.93 & ~34.8 & 6.75 & ~42.5 & 7.16 \\
Fe{\sc ii} & 6516.08 & 2.89 & $-$3.38 & ~39.2 & 7.21 &       &      & ~67.2 & 7.56 & ~42.7 & 6.95 & ~47.8 & 7.23 \\
Ni{\sc i} & 5847.01 & 1.68 & $-$3.44 & ~79.3 & 5.99 & 108.3 & 5.99 & 131.8 & 6.49 & 104.5 & 6.09 & 110.4 & 6.28 \\
Ni{\sc i} & 5996.74 & 4.24 & $-$1.06 & ~22.7 & 5.70 & ~34.9 & 5.90 & ~31.1 & 5.82 & ~31.4 & 5.82 & ~41.8 & 6.09 \\
Ni{\sc i} & 6053.69 & 4.24 & $-$1.07 & ~31.5 & 5.97 & ~53.6 & 6.28 & ~40.5 & 6.03 & ~46.0 & 6.14 &       &      \\
Ni{\sc i} & 6086.29 & 4.27 & $-$0.47 & ~46.2 & 5.82 & ~61.6 & 5.87 & ~47.5 & 5.61 & ~47.6 & 5.61 & ~68.0 & 6.10 \\
Ni{\sc i} & 6108.12 & 1.68 & $-$2.49 & 117.7 & 5.92 & 174.9 & 6.12 & 157.1 & 5.90 & 147.9 & 5.90 & 156.2 & 6.11 \\
Ni{\sc i} & 6111.08 & 4.09 & $-$0.83 & ~55.4 & 6.17 & ~72.4 & 6.17 & ~53.6 & 5.84 & ~61.5 & 6.00 & ~63.5 & 6.10 \\
Ni{\sc i} & 6128.98 & 1.68 & $-$3.39 & ~76.7 & 5.80 & 116.0 & 6.02 & 110.6 & 5.96 & 101.8 & 5.93 & 111.2 & 6.15 \\
Ni{\sc i} & 6130.14 & 4.27 & $-$0.98 & ~30.4 & 5.90 & ~49.1 & 6.15 & ~38.0 & 5.93 & ~39.4 & 5.96 & ~50.1 & 6.23 \\
Ni{\sc i} & 6176.82 & 4.09 & $-$0.26 & ~77.0 & 6.14 & ~81.3 & 5.76 & ~87.3 & 5.88 & ~88.3 & 5.95 & ~83.8 & 5.96 \\
Ni{\sc i} & 6177.25 & 1.83 & $-$3.60 & ~63.4 & 5.90 & ~88.4 & 5.98 & ~89.7 & 6.02 & ~90.6 & 6.14 & ~82.7 & 5.99 \\
Ni{\sc i} & 6186.72 & 4.11 & $-$0.90 & ~50.4 & 6.14 & ~74.2 & 6.30 & ~46.7 & 5.80 & ~65.6 & 6.17 & ~60.0 & 6.13 \\
Ni{\sc i} & 6204.61 & 4.09 & $-$1.15 & ~45.2 & 6.22 & ~66.3 & 6.38 & ~54.8 & 6.18 & ~50.8 & 6.11 & ~61.5 & 6.38 \\
Ni{\sc i} & 6223.99 & 4.11 & $-$0.97 & ~46.0 & 6.09 & ~49.6 & 5.93 & ~39.8 & 5.74 & ~53.1 & 6.00 & ~37.9 & 5.73 \\
Ni{\sc i} & 6230.10 & 4.11 & $-$1.20 & ~43.5 & 6.26 & ~57.9 & 6.31 & ~53.3 & 6.23 & ~56.0 & 6.29 & ~51.0 & 6.24 \\
Ni{\sc i} & 6322.17 & 4.15 & $-$1.21 &       &      & ~45.4 & 6.14 & ~20.6 & 5.59 & ~30.5 & 5.83 & ~43.3 & 6.14 \\
Ni{\sc i} & 6327.60 & 1.68 & $-$3.09 &       &      &       &      &       &      & 128.0 & 6.09 & 138.8 & 6.35 \\
Ni{\sc i} & 6378.26 & 4.15 & $-$0.82 & ~49.4 & 6.08 & ~73.9 & 6.27 & ~51.4 & 5.86 & ~50.3 & 5.85 & ~70.3 & 6.32 \\
Ni{\sc i} & 6384.67 & 4.15 & $-$1.00 & ~46.1 & 6.17 & ~69.4 & 6.36 & ~44.8 & 5.92 & ~36.2 & 5.75 & ~61.1 & 6.30 \\
Ni{\sc i} & 6482.81 & 1.93 & $-$2.85 & ~93.0 & 5.86 & 148.4 & 6.24 & 136.9 & 6.11 & 122.4 & 6.00 & 126.1 & 6.10 \\
Ni{\sc i} & 6532.88 & 1.93 & $-$3.42 &       &      &       &      &       &      & ~95.0 & 6.14 & ~94.2 & 6.14 \\
Ni{\sc i} & 6586.32 & 1.95 & $-$2.79 & ~99.6 & 6.12 & 142.1 & 6.21 & 125.1 & 5.96 & 108.3 & 5.78 & 126.0 & 6.17 \\
Ni{\sc i} & 6598.61 & 4.24 & $-$0.93 &       &      & ~59.0 & 6.23 & ~36.3 & 5.80 & ~42.8 & 5.93 & ~55.1 & 6.23 \\
Ni{\sc i} & 6635.14 & 4.42 & $-$0.75 & ~46.3 & 6.30 & ~57.9 & 6.28 & ~34.4 & 5.82 & ~27.1 & 5.64 & ~52.2 & 6.24 \\
Ni{\sc i} & 6767.78 & 1.83 & $-$2.11 & 131.8 & 5.87 & 189.5 & 6.01 & 165.8 & 5.74 & 148.9 & 5.61 & 173.1 & 6.05 \\
Ni{\sc i} & 6772.32 & 3.66 & $-$1.01 & ~71.3 & 6.09 & ~94.8 & 6.09 & ~74.6 & 5.76 & ~64.6 & 5.62 & ~92.7 & 6.22 \\
Y{\sc i} & 6435.02 & 0.07 & $-$0.82 & ~54.0 & 1.05 & ~98.9 & 1.50 & 112.1 & 1.77 & ~62.6 & 1.23 & ~90.2 & 1.47 \\
Zr{\sc i} & 6127.46 & 0.15 & $-$1.06 & ~86.5 & 2.15 & 121.9 & 2.08 & 116.3 & 2.05 & ~96.8 & 1.97 & ~84.9 & 1.58 \\
Zr{\sc i} & 6134.57 & 0.00 & $-$1.28 & ~67.2 & 1.59 & ~99.4 & 1.70 & 104.3 & 1.84 & ~91.8 & 1.87 & ~87.6 & 1.62 \\
Zr{\sc i} & 6140.46 & 0.52 & $-$1.41 & ~36.8 & 1.57 & ~45.1 & 1.85 & ~41.5 & 1.83 & ~26.8 & 1.73 & ~29.7 & 1.54 \\
Zr{\sc i} & 6143.18 & 0.07 & $-$1.10 & ~81.6 & 1.94 & 121.0 & 1.98 & 106.6 & 1.80 & ~93.5 & 1.83 & ~81.8 & 1.44 \\
Zr{\sc i} & 6445.72 & 1.00 & $-$0.83 &       &      & ~21.9 & 1.56 & ~18.3 & 1.49 & ~12.1 & 1.43 & ~15.4 & 1.31 \\
Ba{\sc ii} & 5853.69 & 0.60 & $-$1.00 & 111.5 & 2.09 & 148.4 & 1.84 & 162.9 & 2.18 & 125.2 & 1.69 & 140.5 & 2.09 \\
Ba{\sc ii} & 6141.75 & 0.70 &  ~~0.00 & 159.2 & 1.89 & 220.6 & 1.93 & 205.8 & 1.82 & 180.3 & 1.68 & 182.9 & 1.74 \\
Ba{\sc ii} & 6496.91 & 0.60 & $-$0.38 & 182.2 & 2.32 & 221.6 & 2.14 & 205.3 & 1.99 & 177.4 & 1.83 & 196.3 & 2.10 \\
Eu{\sc ii} & 6437.64 & 1.32 & $-$0.28 &       &      & ~34.0 & 0.65 & ~22.5 & 0.40 & ~25.1 & 0.50 & ~24.6 & 0.42 \\
Eu{\sc ii} & 6645.11 & 1.38 &  ~~0.20 & ~39.3 & 0.55 & ~54.6 & 0.61 & ~48.6 & 0.52 & ~31.0 & 0.23 & ~41.0 & 0.38 \\
\hline
\end{tabular}
\normalsize
\end{table*}

\begin{table*}
\caption{Equivalent Widths from UVES spectra for non members (in electronic form)}
\scriptsize
\begin{tabular}{lccccccccccccccccc}
\hline
El. &Wavel. &E.P.&log gf& 3734&    & 4360&    & 5308&    & 5341&    & 4329&    & 4453&    & 3092&     \\
    &       &    &      &EW   &log A& EW   &log A& EW   &log A& EW   &log A&EW    &log A& EW   &log A&EW    &log A\\
    &(\AA)  &(eV)&      &(m\AA)&     &(m\AA)&     &(m\AA)&     &(m\AA)&      &(m\AA)&    &(m\AA)&      &(m\AA)&     \\
\hline
O{\sc i}&6300.31&0.00&-9.75& 86.2&8.83& 63.7&8.82& 83.6&8.98& 77.0&8.89&     &    & 69.2&8.78& 95.7&8.89 \\
O{\sc i}&6363.79&0.02&-10.25&44.0&8.82& 39.2&9.03& 29.7&8.77&     &    &     &    & 39.3&8.92& 50.0&8.78 \\
Na{\sc i}&6154.23&2.10&-1.57& 91.1&6.19&166.3&7.22&100.9&6.42&125.3&6.70&108.9&6.62&113.2&6.50& 91.9&6.26 \\
Na{\sc i}&6160.75&2.10&-1.26& 98.9&6.00&169.6&6.96& 95.5&6.03&136.0&6.55&129.6&6.63&120.4&6.30&129.6&6.60 \\
Mg{\sc i}&6318.71&5.11&-1.94& 81.4&7.64&132.4&8.40& 80.8&7.63& 94.4&7.88& 89.7&7.82& 74.2&7.48& 72.1&7.63 \\
Mg{\sc i}&6319.24&5.11&-2.16& 57.6&7.44&102.6&8.16& 39.2&7.14& 70.4&7.71& 80.9&7.88& 61.1&7.49& 64.2&7.69 \\
Al{\sc i}&6696.03&3.14&-1.32& 88.7&6.05&144.7&6.88&118.1&6.66&115.0&6.46&103.6&6.46&118.1&6.52& 98.2&6.28 \\
Al{\sc i}&6698.67&3.14&-1.62& 59.2&5.87&109.9&6.64& 99.3&6.64& 77.1&6.14& 64.1&6.09& 79.3&6.21& 69.8&6.02 \\
Si{\sc i}&5948.55&5.08&-1.23& 81.3&7.55&121.7&8.18& 62.6&7.13& 85.1&7.68&104.5&7.85& 93.4&7.67& 58.2&7.33 \\
Si{\sc i}&6125.03&5.62&-1.57&     &    & 84.3&8.55&     &    &     &    & 48.0&7.76&     &    &     &     \\
Si{\sc i}&6145.02&5.62&-1.49& 45.2&7.77& 87.6&8.51& 37.5&7.53& 25.7&7.38& 50.7&7.74& 36.0&7.51& 28.7&7.55 \\
Ca{\sc i}&5857.46&2.93& 0.24&168.3&5.77&213.9&6.25&165.6&5.86&195.1&6.02&201.9&6.32&204.5&6.14&128.0&5.34 \\
Ca{\sc i}&5867.57&2.93&-1.49& 58.8&5.65& 97.1&6.30& 70.3&6.01&105.2&6.49& 83.3&6.31& 79.1&6.04& 80.0&6.12 \\
Ca{\sc i}&6161.30&2.52&-1.27&129.5&6.16&160.8&6.56&120.8&6.16&147.7&6.40&     &    &142.2&6.32&110.3&5.93 \\
Ca{\sc i}&6166.44&2.52&-1.14&121.0&5.88&163.0&6.47&124.7&6.10&     &    &     &    &     &    &     &     \\
Ca{\sc i}&6169.04&2.52&-0.80&159.5&6.14&192.2&6.52&     &    &166.6&6.20&162.4&6.44&158.7&6.12&121.5&5.68 \\
Ca{\sc i}&6169.56&2.52&-0.48&159.4&5.82&196.9&6.25&     &    &195.4&6.20&179.2&6.32&167.5&5.92&149.2&5.74 \\
Ca{\sc i}&6439.08&2.52& 0.39&211.3&5.64&243.6&5.93&211.9&5.78&223.3&5.72&222.4&6.00&229.5&5.82&185.4&5.41 \\
Ca{\sc i}&6449.82&2.52&-0.50&171.2&6.10&193.2&6.29&175.1&6.30&196.3&6.35&177.3&6.44&190.8&6.30&     &     \\
Ca{\sc i}&6455.60&2.52&-1.29&106.4&5.75&141.0&6.24&122.2&6.22&123.4&5.99&118.6&6.21&128.8&6.10&101.4&5.77 \\
Ca{\sc i}&6462.57&2.52& 0.26&271.5&6.16&315.6&6.50&272.3&6.31&274.2&6.20&262.9&6.39&281.3&6.30&226.1&5.84 \\
Ca{\sc i}&6471.67&2.52&-0.69&148.0&5.94&185.6&6.38&156.0&6.22&     &    &155.6&6.30&157.2&6.02&125.8&5.71 \\
Ca{\sc i}&6493.79&2.52&-0.11&185.7&5.90&217.3&6.19&188.0&6.06&186.6&5.86&177.3&6.05&191.6&5.93&154.3&5.57 \\
Ca{\sc i}&6499.65&2.52&-0.82&145.2&6.02&177.6&6.39&141.7&6.11&151.9&6.06&147.0&6.28&     &    &     &     \\
Sc{\sc ii}&6245.62&1.51&-1.05&     &    & 99.5&3.24& 91.3&3.15&105.6&3.44& 89.9&3.21& 87.9&2.96& 56.1&2.45 \\
Sc{\sc ii}&6279.74&1.50&-1.16& 72.6&2.80& 82.0&3.06& 97.2&3.41&109.9&3.65& 71.5&2.98& 85.1&3.04& 59.5&2.66 \\
Sc{\sc ii}&6604.60&1.36&-1.15& 94.5&2.98&133.5&3.70&     &    & 76.6&2.74& 71.8&2.73&116.6&3.36& 84.0&2.96 \\
Ti{\sc i}&5866.46&1.07&-0.84&162.8&4.86&200.9&5.40&164.7&5.14&183.9&5.16&165.9&5.31&186.7&5.25&     &     \\
Ti{\sc i}&5922.12&1.05&-1.47&143.5&5.03&     &    &141.3&5.25&162.8&5.30&137.1&5.30&148.1&5.06&110.1&4.42 \\
Ti{\sc i}&5978.55&1.87&-0.50&114.8&4.72&144.0&5.14&127.1&5.20&147.9&5.29&124.3&5.24&130.4&4.97& 95.3&4.36 \\
Ti{\sc i}&6091.18&2.27&-0.42&111.4&5.16&132.3&5.42& 94.4&4.96&119.1&5.24& 90.2&4.95&104.9&4.96& 96.4&4.93 \\
Ti{\sc i}&6126.22&1.07&-1.42&147.0&5.02&159.8&5.11&145.5&5.24&160.0&5.17&122.2&4.90&157.8&5.15&129.0&4.84 \\
Ti{\sc i}&6258.11&1.44&-0.36&159.9&4.71&190.3&5.12&159.8&4.94&198.4&5.24&     &    &172.6&4.89&128.5&4.32 \\
Ti{\sc i}&6554.24&1.44&-1.22&114.6&4.62&155.8&5.26&112.5&4.82&     &    &127.5&5.21&138.2&5.02&     &     \\
Ti{\sc ii}&6606.98&2.06&-2.90& 24.5&4.75& 66.0&5.72& 28.6&4.92& 36.5&5.19& 38.1&5.21& 63.7&5.60& 30.4&4.95 \\
V{\sc i}&6002.31&1.22&-1.77& 38.7&3.51& 89.0&4.39& 52.1&3.98& 72.3&4.08& 45.6&3.96& 56.6&3.94& 71.4&4.03 \\
V{\sc i}&6039.73&1.06&-0.65&110.9&3.46&159.9&4.26&109.1&3.67&140.2&3.99&117.0&3.96&132.6&3.85&117.3&3.79 \\
V{\sc i}&6081.45&1.05&-0.58&126.1&3.80&163.2&4.38&     &    &159.7&4.36&130.6&4.26&157.4&4.32&128.9&4.03 \\
V{\sc i}&6090.22&1.08&-0.06&148.7&3.76&183.2&4.27&138.8&3.83&177.6&4.19&138.8&3.97&167.2&4.04&147.9&3.87 \\
V{\sc i}&6111.65&1.04&-0.71&     &    &     &    &135.9&4.20&174.0&4.54&136.8&4.35&147.7&4.08&     &     \\
V{\sc i}&6119.53&1.06&-0.32&134.1&3.71&185.3&4.50&128.1&3.83&162.8&4.16&125.4&3.90&147.0&3.88&131.3&3.82 \\
V{\sc i}&6135.38&1.05&-0.75&119.8&3.70&166.4&4.39&118.3&3.91&145.9&4.09&116.9&3.98&135.2&3.92&108.0&3.56 \\
V{\sc i}&6199.19&0.29&-1.28&170.3&3.94&202.1&4.42&172.6&4.26&211.8&4.56&171.7&4.39&202.6&4.48&159.5&3.85 \\
V{\sc i}&6216.36&0.28&-0.81&179.7&3.58&211.2&4.04&172.3&3.74&215.9&4.09&171.6&3.88&198.1&3.88&     &     \\
V{\sc i}&6242.81&0.26&-1.55&     &    &174.1&4.15&138.7&3.94&     &    &     &    &     &    &     &     \\
V{\sc i}&6251.82&0.29&-1.34&     &    &     &    &158.7&4.14&190.8&4.38&147.0&4.07&188.2&4.37&129.1&3.40 \\
V{\sc i}&6256.90&0.28&-2.01&125.6&3.90&166.3&4.50&112.5&3.91&145.6&4.18&114.8&4.06&137.9&4.08&120.0&3.93 \\
V{\sc i}&6274.65&0.27&-1.67&     &    &     &    &132.2&3.97&163.0&4.14&125.1&3.94&153.5&3.99&118.5&3.54 \\
V{\sc i}&6285.17&0.28&-1.51&     &    &     &    &134.4&3.86&175.8&4.25&146.9&4.22&166.0&4.08&136.8&3.83 \\
V{\sc i}&6292.83&0.29&-1.47&     &    &     &    &155.4&4.21&207.2&4.66&137.8&4.03&     &    &     &     \\
V{\sc i}&6531.43&1.22&-0.84& 97.3&3.48&169.1&4.62&107.0&3.91&127.9&4.02&     &    &122.9&3.94&105.5&3.77 \\
\hline
\end{tabular}
\normalsize
\label{t:ewidths2}
\end{table*}

\addtocounter{table}{-1}

\begin{table*}
\caption{Continued}
\scriptsize
\begin{tabular}{lccccccccccccccccc}
\hline
El. &Wavel. &E.P.&log gf& 3734&    & 4360&    & 5308&    & 5341&    & 4329&    & 4453&    & 3092&     \\
    &       &    &      &EW   &log A& EW   &log A& EW   &log A& EW   &log A&EW    &log A& EW   &log A&EW    &log A\\
    &(\AA)  &(eV)&      &(m\AA)&     &(m\AA)&     &(m\AA)&     &(m\AA)&      &(m\AA)&    &(m\AA)&      &(m\AA)&     \\
\hline
Cr{\sc i}&6330.10&0.94&-2.87&121.9&5.04&180.7&5.98&131.6&5.46&155.2&5.60&     &    &149.4&5.48&150.2&5.72 \\
Mn{\sc i}&6013.50&3.07&-0.25&144.8&5.14&203.3&5.84&139.8&5.16&182.6&5.61&183.0&5.81&178.5&5.53&121.1&4.98 \\
Mn{\sc i}&6016.65&3.07&-0.09&155.2&5.18&192.8&5.66&141.4&5.10&172.4&5.43&168.0&5.58&160.1&5.23&113.6&4.72 \\
Mn{\sc i}&6021.80&3.08& 0.03&138.3&4.96&215.1&5.85&136.4&5.02&174.3&5.44&166.6&5.55&171.0&5.36&121.9&4.89 \\
Fe{\sc i}&5835.11&4.26&-2.18& 35.5&7.23& 76.3&8.30&     &    &     &    &     &    & 63.2&7.80&     &     \\
Fe{\sc i}&5855.09&4.61&-1.48& 43.7&7.19&     &    & 49.8&7.35& 49.0&7.36& 60.8&7.64& 54.9&7.40& 37.4&7.17 \\
Fe{\sc i}&5856.10&4.29&-1.57& 69.1&7.42& 95.5&7.90& 60.3&7.25& 88.6&7.89& 81.1&7.80& 81.7&7.60&     &     \\
Fe{\sc i}&5858.78&4.22&-2.19& 29.7&7.05&     &    & 33.5&7.50& 42.5&7.42& 46.3&7.52& 42.7&7.36&     &     \\
Fe{\sc i}&5859.60&4.55&-0.70& 84.0&7.20&     &    & 81.6&7.16&114.2&7.80&115.7&7.94&111.4&7.63& 73.6&7.25 \\
Fe{\sc i}&5862.37&4.55&-0.50& 99.1&7.31&116.3&7.54& 81.3&6.95&110.4&7.54&124.4&7.88&104.0&7.30& 84.3&7.28 \\
Fe{\sc i}&5881.28&4.61&-1.76& 39.3&7.36&     &    & 47.8&7.58& 68.4&8.04&     &    & 59.6&7.76&     &     \\
Fe{\sc i}&5902.48&4.59&-1.86&     &    &     &    & 49.9&7.71& 41.1&7.55& 53.7&7.84& 43.3&7.52&     &     \\
Fe{\sc i}&5905.68&4.65&-0.76& 81.7&7.38&124.1&8.10& 81.3&7.37&     &    &     &    &     &    &     &     \\
Fe{\sc i}&5927.80&4.65&-1.06& 63.4&7.26& 87.3&7.69& 58.0&7.15& 81.0&7.67& 78.4&7.66& 65.1&7.22& 49.1&7.14 \\
Fe{\sc i}&5929.68&4.55&-1.24& 73.5&7.53&100.5&8.00& 51.0&7.05& 86.4&7.83& 79.6&7.74& 73.8&7.44& 49.0&7.17 \\
Fe{\sc i}&5930.19&4.65&-0.29&     &    &150.8&8.06& 91.0&7.11&125.9&7.77&125.9&7.88&118.9&7.52& 92.2&7.45 \\
Fe{\sc i}&5934.67&3.93&-1.15& 98.1&7.12&131.1&7.65&123.5&7.65&145.4&7.96&134.5&7.94&129.3&7.59&101.9&7.50 \\
Fe{\sc i}&5956.71&0.86&-4.60&159.0&7.41&     &    &147.5&7.41&163.0&7.52&143.9&7.51&129.3&7.58&     &     \\
Fe{\sc i}&5976.79&3.94&-1.33& 88.1&7.10&     &    & 84.9&7.07&119.7&7.76&114.7&7.82&111.6&7.48& 68.5&6.92 \\
Fe{\sc i}&5984.83&4.73&-0.39&103.6&7.48&     &    & 89.5&7.22&135.1&7.98&     &    &     &    & 99.2&7.65 \\
Fe{\sc i}&6003.02&3.88&-1.08&110.3&7.24&142.9&7.72&107.3&7.22&135.2&7.68&134.9&7.82&127.7&7.44&107.9&7.47 \\
Fe{\sc i}&6007.97&4.65&-0.82& 67.8&7.09&101.9&7.70& 73.9&7.24& 81.0&7.40& 97.6&7.82& 84.4&7.35&     &     \\
Fe{\sc i}&6008.57&3.88&-0.96&109.9&7.12&146.9&7.68&102.8&7.02&129.6&7.48&134.2&7.70&130.5&7.38& 88.6&6.97 \\
Fe{\sc i}&6027.06&4.08&-1.23&104.6&7.55&118.0&7.69& 97.7&7.44&114.9&7.76&101.9&7.62&102.7&7.39& 88.2&7.50 \\
Fe{\sc i}&6056.01&4.73&-0.42& 86.8&7.25&111.2&7.62& 88.0&7.28&102.6&7.58&104.2&7.71& 92.7&7.25&     &     \\
Fe{\sc i}&6078.50&4.80&-0.48&     &    &127.8&7.99&100.5&7.60&110.9&7.80&108.6&7.85& 99.9&7.49& 71.0&7.28 \\
Fe{\sc i}&6079.02&4.65&-0.95& 76.9&7.45&104.6&7.92& 66.5&7.23& 89.0&7.73&     &    & 79.1&7.40& 68.5&7.55 \\
Fe{\sc i}&6082.72&2.22&-3.57&112.7&7.50&     &    &119.4&7.77&135.0&7.96&116.1&7.82&123.3&7.60&     &     \\
Fe{\sc i}&6089.57&4.58&-1.28&     &    &     &    &     &    & 94.0&8.03& 85.6&7.93& 93.8&7.90&     &     \\
Fe{\sc i}&6093.65&4.61&-1.32& 64.5&7.49&     &    & 60.6&7.42& 59.0&7.40& 61.4&7.49& 61.5&7.36& 58.2&7.59 \\
Fe{\sc i}&6094.38&4.65&-1.56& 43.4&7.31&     &    & 40.2&7.27& 41.1&7.32& 48.0&7.48& 45.2&7.34&     &     \\
Fe{\sc i}&6096.67&3.98&-1.77& 73.2&7.26&109.2&7.93& 78.1&7.41& 86.0&7.56& 85.1&7.64& 79.0&7.30& 54.1&7.02 \\
Fe{\sc i}&6098.25&4.56&-1.81& 47.1&7.52&     &    & 46.4&7.54& 58.5&7.82& 55.3&7.78& 55.9&7.68&     &     \\
Fe{\sc i}&6137.00&2.20&-2.95&135.3&7.26&157.4&7.56&146.0&7.56&164.4&7.77&147.8&7.73&170.5&7.77&     &     \\
Fe{\sc i}&6151.62&2.18&-3.30&135.2&7.59&148.8&7.72&118.7&7.38&123.3&7.34&121.7&7.56&137.1&7.50&103.0&7.21 \\
Fe{\sc i}&6165.36&4.14&-1.50& 70.1&7.19& 96.8&7.60& 63.1&7.02& 88.7&7.56& 84.5&7.55& 77.0&7.20& 50.9&6.89 \\
Fe{\sc i}&6173.34&2.22&-2.88&147.7&7.44&184.9&7.94&     &    &165.7&7.74&144.0&7.62&154.8&7.47&126.0&7.35 \\
Fe{\sc i}&6188.00&3.94&-1.63& 81.0&7.22&     &    & 76.9&7.18& 97.6&7.59& 97.8&7.73& 81.2&7.15& 59.2&6.96 \\
Fe{\sc i}&6200.32&2.61&-2.44&151.2&7.59&160.4&7.66&121.6&7.18&159.4&7.74&138.9&7.62&149.2&7.47&     &     \\
Fe{\sc i}&6226.74&3.88&-2.08& 73.2&7.42& 79.1&7.50& 59.1&7.17& 66.0&7.30& 72.0&7.51& 61.0&7.13& 57.0&7.26 \\
Fe{\sc i}&6232.65&3.65&-1.22&140.0&7.52&179.2&8.03&123.7&7.34&141.8&7.59&131.6&7.58&135.4&7.38&     &     \\
Fe{\sc i}&6240.65&2.22&-3.23&128.9&7.45&148.5&7.69&108.1&7.14&121.5&7.29&     &    &133.3&7.42&101.3&7.15 \\
Fe{\sc i}&6246.33&3.60&-0.73&140.4&6.98&     &    &136.2&6.97&     &    &     &    &     &    &     &     \\
Fe{\sc i}&6265.14&2.18&-2.55&169.2&7.35&     &    &168.6&7.47&     &    &     &    &     &    &147.5&7.27 \\
Fe{\sc i}&6270.23&2.86&-2.46&     &    &137.1&7.60&104.8&7.19&129.8&7.60&125.4&7.74&122.7&7.34& 90.9&7.07 \\
Fe{\sc i}&6297.80&2.22&-2.74&     &    &     &    &155.0&7.52&168.2&7.64&148.6&7.55&170.8&7.59&128.7&7.26 \\
Fe{\sc i}&6301.51&3.65&-0.72&142.6&7.04&     &    &162.7&7.32&152.3&7.20&160.1&7.44&     &    &     &     \\
Fe{\sc i}&6311.50&2.83&-3.16&101.5&7.65&121.2&7.92& 96.7&7.62&116.2&7.94& 95.3&7.68&108.4&7.67& 93.2&7.76 \\
Fe{\sc i}&6315.81&4.08&-1.68& 78.0&7.39& 91.8&7.59& 77.5&7.41& 90.9&7.68& 70.4&7.32& 82.6&7.40& 78.1&7.67 \\
Fe{\sc i}&6322.69&2.59&-2.43&140.6&7.34&175.0&7.80&148.4&7.56&151.2&7.53&147.3&7.68&156.3&7.50&120.5&7.28 \\
Fe{\sc i}&6330.85&4.73&-1.22& 49.2&7.20&     &    & 61.4&7.47& 71.1&7.70& 70.6&7.73& 48.3&7.15&     &     \\
Fe{\sc i}&6380.75&4.19&-1.37& 80.0&7.28&     &    & 96.6&7.65&103.2&7.77&101.8&7.85&102.1&7.62& 76.8&7.48 \\
Fe{\sc i}&6392.54&2.28&-3.97& 79.6&7.16&111.8&7.76& 83.4&7.36& 84.4&7.30& 82.7&7.44& 95.3&7.44& 77.7&7.34 \\
Fe{\sc i}&6411.66&3.65&-0.60&143.2&6.92&     &    &155.4&7.12&     &    &     &    &167.8&7.18&     &     \\
Fe{\sc i}&6481.88&2.28&-2.98&136.0&7.36&179.4&7.99&150.3&7.71&152.5&7.67&137.5&7.64&160.1&7.68&125.3&7.47 \\
Fe{\sc i}&6498.94&0.96&-4.70&147.3&7.35&198.5&8.10&160.0&7.74&174.3&7.83&     &    &184.5&7.89&139.2&7.47 \\
\hline
\end{tabular}
\normalsize
\end{table*}

\addtocounter{table}{-1}

\begin{table*}
\caption{Continued}
\scriptsize
\begin{tabular}{lccccccccccccccccc}
\hline
El. &Wavel. &E.P.&log gf& 3734&    & 4360&    & 5308&    & 5341&    & 4329&    & 4453&    & 3092&     \\
    &       &    &      &EW   &log A& EW   &log A& EW   &log A& EW   &log A&EW    &log A& EW   &log A&EW    &log A\\
    &(\AA)  &(eV)&      &(m\AA)&     &(m\AA)&     &(m\AA)&     &(m\AA)&      &(m\AA)&    &(m\AA)&      &(m\AA)&     \\
\hline
Fe{\sc i}&6518.37&2.83&-2.46&108.5&7.06&140.6&7.54&     &    &     &    &112.7&7.33&127.0&7.30& 93.5&7.04 \\
Fe{\sc i}&6533.94&4.56&-1.29& 59.2&7.24& 92.1&7.84&     &    & 84.9&7.82& 78.5&7.73&     &    &     &     \\
Fe{\sc i}&6574.25&0.99&-5.00&125.2&7.23&165.7&7.89&122.7&7.35&136.4&7.46&121.8&7.46&147.8&7.58&109.9&7.20 \\
Fe{\sc i}&6581.22&1.49&-4.68&     &    &123.0&7.52&129.4&7.87&125.5&7.67&116.3&7.74&139.9&7.83&111.0&7.66 \\
Fe{\sc i}&6593.88&2.43&-2.42&144.2&7.11&     &    &160.5&7.47&155.8&7.33&158.4&7.58&175.3&7.50&133.0&7.20 \\
Fe{\sc i}&6608.04&2.28&-3.96& 82.0&7.18&131.2&8.08& 94.6&7.56& 97.3&7.53& 96.1&7.69& 92.2&7.35& 83.6&7.45 \\
Fe{\sc i}&6609.12&2.56&-2.69&130.9&7.34&174.5&7.95&134.6&7.50&149.3&7.68&141.6&7.75&164.7&7.80&108.0&7.21 \\
Fe{\sc i}&6627.56&4.55&-1.50& 40.2&7.03&     &    & 55.4&7.38& 55.7&7.42& 59.5&7.52& 82.7&7.85& 35.4&7.04 \\
Fe{\sc i}&6633.76&4.56&-0.82& 83.2&7.30&119.5&7.88&100.0&7.65& 88.8&7.43&109.4&7.92&107.6&7.66& 77.3&7.48 \\
Fe{\sc i}&6703.58&2.76&-3.01& 87.7&7.02&129.3&7.75&101.6&7.42&102.1&7.34&100.8&7.49&119.4&7.57& 83.6&7.20 \\
Fe{\sc i}&6713.75&4.80&-1.41& 42.5&7.33&     &    & 45.9&7.41& 30.8&7.13& 38.1&7.28& 56.2&7.57&     &     \\
Fe{\sc i}&6725.36&4.10&-2.21&     &    & 70.1&7.72& 50.0&7.37& 40.4&7.20& 60.7&7.64& 64.0&7.58& 34.4&7.08 \\
Fe{\sc i}&6726.67&4.61&-1.10& 57.2&7.07& 82.4&7.52& 69.7&7.34&     &    & 77.2&7.55& 75.7&7.37& 37.8&6.78 \\
Fe{\sc i}&6733.15&4.64&-1.44& 32.4&6.91& 79.3&7.85& 54.1&7.40& 61.2&7.59& 54.6&7.46& 80.2&7.86& 53.6&7.60 \\
Fe{\sc i}&6750.16&2.42&-2.62&136.3&7.11&167.2&7.53&160.8&7.61&157.2&7.49&150.9&7.59&169.0&7.56&123.2&7.18 \\
Fe{\sc i}&6786.86&4.19&-1.90& 41.9&6.97& 85.2&7.80& 61.8&7.42& 49.3&7.19& 61.9&7.47& 78.4&7.66& 48.7&7.28 \\
Fe{\sc ii}&5991.38&3.15&-3.56& 31.0&7.31&     &    & 46.0&7.67&     &    &     &    & 37.0&7.42&     &     \\
Fe{\sc ii}&6084.10&3.20&-3.80&     &    &     &    &     &    & 30.2&7.75& 35.0&7.70& 35.2&7.67&     &     \\
Fe{\sc ii}&6113.33&3.22&-4.13&     &    &     &    & 27.1&7.78&     &    & 23.8&7.70&     &    &     &     \\
Fe{\sc ii}&6149.25&3.89&-2.73&     &    &     &    & 49.1&7.86& 21.1&7.22& 27.1&7.21& 29.2&7.27&     &     \\
Fe{\sc ii}&6247.56&3.89&-2.33&     &    & 52.5&7.64& 54.3&7.60& 54.3&7.87& 43.2&7.33& 41.5&7.24& 27.6&7.27 \\
Fe{\sc ii}&6369.46&2.89&-4.21& 25.5&7.47& 28.0&7.64& 32.1&7.62&     &    & 28.7&7.55& 27.1&7.48&     &     \\
Fe{\sc ii}&6416.93&3.89&-2.70& 33.4&7.49&     &    &     &    & 29.1&7.50&     &    &     &    &     &     \\
Fe{\sc ii}&6432.68&2.89&-3.58& 35.3&7.15& 65.4&7.93& 56.4&7.64& 30.6&7.18& 38.9&7.22& 45.0&7.34& 24.1&7.07 \\
Fe{\sc ii}&6456.39&3.90&-2.10& 39.5&7.10& 75.6&7.97& 58.3&7.50& 47.6&7.48& 44.7&7.16& 55.6&7.40& 21.8&6.82 \\
Fe{\sc ii}&6516.08&2.89&-3.38& 53.5&7.45&     &    & 63.9&7.62&     &    &     &    &     &    &     &     \\
Ni{\sc i}&5847.01&1.68&-3.44&102.3&6.39&132.0&6.92&104.1&6.55&113.0&6.67&     &    &     &    & 90.3&6.39 \\
Ni{\sc i}&5996.74&4.24&-1.06& 38.9&6.10& 67.6&6.72& 39.1&6.12& 42.6&6.27& 54.0&6.51& 47.8&6.28& 45.0&6.44 \\
Ni{\sc i}&6053.69&4.24&-1.07& 37.5&6.08& 84.3&7.06& 28.8&5.88& 64.7&6.76& 55.7&6.56& 48.9&6.31&     &     \\
Ni{\sc i}&6086.29&4.27&-0.47& 52.5&5.87& 98.2&6.78& 65.2&6.16& 88.0&6.70& 77.5&6.50& 73.0&6.24& 55.2&6.18 \\
Ni{\sc i}&6108.12&1.68&-2.49&139.7&6.13&178.3&6.70&134.8&6.17&154.1&6.41&136.9&6.34&162.0&6.43&124.7&6.17 \\
Ni{\sc i}&6111.08&4.09&-0.83& 65.4&6.28& 95.4&6.82& 66.2&6.30& 65.1&6.31& 73.1&6.52& 71.1&6.32& 49.1&6.11 \\
Ni{\sc i}&6128.98&1.68&-3.39&107.8&6.41&132.7&6.80& 89.7&6.10&108.0&6.42&107.6&6.63&110.8&6.37& 81.0&6.02 \\
Ni{\sc i}&6130.14&4.27&-0.98& 32.7&5.92& 77.0&6.86& 34.3&5.97& 46.6&6.31& 47.0&6.31& 52.4&6.33& 37.8&6.20 \\
Ni{\sc i}&6176.82&4.09&-0.26& 91.9&6.29&110.0&6.53& 87.8&6.22& 93.6&6.34& 93.2&6.41& 91.9&6.17& 51.1&5.60 \\
Ni{\sc i}&6177.25&1.83&-3.60& 80.0&6.19&102.3&6.63& 78.9&6.28&104.3&6.76& 77.8&6.35& 77.1&6.12&     &     \\
Ni{\sc i}&6186.72&4.11&-0.90& 43.0&5.87& 89.7&6.81& 59.4&6.25& 78.0&6.69& 70.6&6.55& 57.7&6.14&     &     \\
Ni{\sc i}&6204.61&4.09&-1.15& 61.4&6.51& 99.9&7.23& 58.7&6.46& 37.4&6.04& 64.5&6.64& 65.7&6.53&     &     \\
Ni{\sc i}&6223.99&4.11&-0.97& 49.9&6.09& 75.0&6.60& 52.2&6.16& 67.6&6.53& 58.8&6.36& 47.0&6.00& 28.0&5.68 \\
Ni{\sc i}&6230.10&4.11&-1.20& 47.1&6.26& 83.9&7.00& 53.6&6.42& 69.0&6.79& 52.6&6.45& 52.3&6.34& 31.7&6.02 \\
Ni{\sc i}&6322.17&4.15&-1.21& 34.8&6.03& 75.6&6.88& 33.8&6.03& 25.5&5.89& 48.3&6.40& 30.9&5.95&     &     \\
Ni{\sc i}&6327.60&1.68&-3.09&117.8&6.28&     &    &121.4&6.47&132.8&6.57&124.1&6.65&142.5&6.64&101.0&6.20 \\
Ni{\sc i}&6378.26&4.15&-0.82& 51.5&6.02&107.7&7.11& 77.0&6.60& 57.6&6.22& 66.7&6.43& 65.8&6.27& 54.1&6.32 \\
Ni{\sc i}&6384.67&4.15&-1.00& 38.6&5.91& 85.5&6.86& 66.5&6.54& 56.7&6.38& 58.4&6.42& 70.4&6.54& 39.5&6.10 \\
Ni{\sc i}&6482.81&1.93&-2.85&110.0&6.08&149.3&6.72&128.4&6.47&129.1&6.45&106.9&6.27&132.4&6.41&103.4&6.14 \\
Ni{\sc i}&6532.88&1.93&-3.42& 81.0&6.12&117.8&6.81& 93.5&6.50&106.5&6.71& 93.1&6.60& 94.2&6.36& 66.4&5.98 \\
Ni{\sc i}&6586.32&1.95&-2.79&108.3&6.12&158.6&6.96&     &    &122.7&6.42&114.9&6.48&133.3&6.50& 92.3&6.05 \\
Ni{\sc i}&6598.61&4.24&-0.93& 31.1&5.78& 81.8&6.83& 59.7&6.43& 48.6&6.26& 50.5&6.28& 67.1&6.52& 52.1&6.50 \\
Ni{\sc i}&6635.14&4.42&-0.75& 32.6&5.88& 80.0&6.86& 51.6&6.32& 34.2&6.00& 51.5&6.36& 46.5&6.17& 29.1&5.92 \\
Ni{\sc i}&6767.78&1.83&-2.11&145.3&5.92&182.6&6.41&157.4&6.22&152.4&6.07&146.0&6.17&168.9&6.20&124.4&5.86 \\
Ni{\sc i}&6772.32&3.66&-1.01& 68.9&5.89&103.1&6.51& 79.4&6.15& 78.0&6.14& 84.5&6.32& 87.3&6.19& 63.9&6.04 \\
Y{\sc i}&6435.02&0.07&-0.82& 61.7&1.30&126.3&2.54& 50.1&1.41& 75.7&1.57& 56.4&1.62& 84.0&1.82& 64.2&1.27 \\
Zr{\sc i}&6127.46&0.15&-1.06& 85.2&2.00&118.9&2.64& 78.8&2.18& 89.6&2.07& 72.4&2.16& 86.7&2.10& 79.6&2.02 \\
Zr{\sc i}&6134.57&0.00&-1.28& 87.0&2.03& 96.2&2.21& 73.7&2.07& 78.0&1.84& 60.3&1.91& 78.0&1.94& 85.7&2.16 \\
Zr{\sc i}&6140.46&0.52&-1.41& 26.7&1.75&     &    &     &    & 25.1&1.79& 39.0&2.40& 42.7&2.26& 38.2&1.75 \\
Zr{\sc i}&6143.18&0.07&-1.10& 74.7&1.69&116.3&2.51& 73.0&1.98& 98.7&2.17& 72.7&2.09& 87.9&2.04& 98.8&2.40 \\
Zr{\sc i}&6445.72&1.00&-0.83& 14.0&1.50& 26.9&2.07&     &    &     &    &     &    &     &    &     &     \\
Ba{\sc ii}&5853.69&0.60&-1.00&128.3&2.28&121.3&1.94&124.3&2.35&122.9&2.16&132.4&2.69&147.2&2.47&102.3&2.10 \\
Ba{\sc ii}&6141.75&0.70& 0.00&185.0&2.08&212.4&2.44&170.5&2.07&197.6&2.30&172.1&2.26&190.1&2.14&150.7&1.86 \\
Ba{\sc ii}&6496.91&0.60&-0.38&     &    &194.3&2.44&166.5&2.24&183.0&2.36&167.1&2.41&173.3&2.10&168.5&2.25 \\
Eu{\sc ii}&6437.64&1.32&-0.28&     &    & 35.7&1.04& 33.4&0.94& 18.6&0.58& 29.9&0.94& 28.9&0.80& 15.2&0.27 \\
Eu{\sc ii}&6645.11&1.38& 0.20& 28.9&0.30& 50.6&0.92& 46.0&0.81& 39.6&0.69& 34.3&0.64& 44.5&0.72& 31.1&0.40 \\
\hline
\end{tabular}
\normalsize
\end{table*}

\section{Metallicities}

\subsection{Equivalent Widths}

The equivalent widths (EWs) were measured on the spectra using the
ROSA code (Gratton 1988; see Table~\ref{t:ewidths} for stars member
of \object{NGC 6441}, and Table~\ref{t:ewidths2} for the field stars) with
Gaussian fittings to the measured profiles: these exploit a linear
relation between EWs and FWHM of the lines, derived from a subset of
lines characterized by cleaner profiles (see Bragaglia et al. 2001 for
further details on this procedure). Since the observed stars span a
very limited atmospheric parameter range, errors in these EWs may be estimated by
comparing EWs measures for individual stars with the EW values for each single line 
averaged over the whole sample. The values listed in Column 9 of 
Table~\ref{t:uvesphot} as errors on the EWs measurements 
are the $r.m.s.$ of residuals around the best fit
line for EWs vs. $<$EWs$>$ (after eliminating a few outliers).
 These errors may be slightly
overestimated, due to real star-to-star differences. They are roughly
reproduced by the formula $\sigma$(EW)$\sim
380/(S/N)$~m\AA. Considering the resolution and sampling of the
spectra, the errors in the EWs are about 4 times larger than
expectations based on photon noise statistics (Cayrel 1988), showing
that errors are dominated by uncertainties in the correct positioning
of the continuum level; the observed errors could be justified by
errors of about 1\% for the best cases and about 2\% for the worst
ones. Due to the problems in background
subtraction in the green-yellow part of the spectra, only lines with
wavelength $>5800$~\AA\ were used.

\begin{table}
\begin{center}
\caption{Atmospheric Parameters for stars observed with UVES}
\begin{tabular}{lcccc}
\hline
Star &T$_{\rm eff}$ &$\log g$ &$[$A/H$]$ &$v_t$      \\
     &     (K)      &         &          &(km s$^{-1}$)\\
\hline
\multicolumn{5}{c}{Stars of \object{NGC 6441}}\\
\hline
7004050 & 3956 & 1.35 & $-$0.40 &0.90 \\
7004434 & 3984 & 1.40 & $-$0.40 &1.75 \\
7004463 & 4000 & 1.41 & $-$0.40 &1.70 \\
7004487 & 4074 & 1.48 & $-$0.40 &1.55 \\
8002961 & 3942 & 1.23 & $-$0.40 &1.40 \\ 
\hline
\multicolumn{5}{c}{Field stars}\\
\hline
6003734 & 4061 & 1.53 & $-$0.28 &1.20 \\ 
6004360 & 4163 & 1.68 &  ~~0.26 &1.55 \\ 
6005308 & 4214 & 1.77 & $-$0.17 &1.20 \\ 
6005341 & 4079 & 1.67 &  ~~0.05 &1.35 \\ 
7004329 & 4257 & 1.73 &  ~~0.11 &1.15 \\ 
7004453 & 4176 & 1.69 & $-$0.05 &1.45 \\ 
8003092 & 3887 & 1.32 & $-$0.32 &0.80 \\
\hline
\end{tabular}
\label{t:uvesatmo}
\end{center}
\end{table}

\subsection{Atmospheric Parameters}

We performed a standard line analysis on the equivalent widths
measured on our spectra, using model atmospheres extracted by
interpolation from the grid by Kurucz (1992; models with the
overshooting option switched off). Atmospheric parameters defining
these model atmospheres were obtained as follows.

Whenever possible, effective temperatures were derived from de-reddened
$V-K$\ colors, obtained by combining our $V$\ magnitudes with $K$\
ones drawn from the 2MASS catalog (Cutri et al. 2003), using the
calibration by Alonso et al. (1999). The
reddening we adopted is $E(B-V)=0.49$, which is the average value
between various literature determinations. Zinn (1980) and Reed et
al. (1988) obtained $E(B-V)=0.47$\ and $E(B-V)=0.49$\ respectively
from integrated photometry; Layden et al. (1999) and Pritzl et
al. (2001) found $E(B-V)=0.45\pm 0.05$\ and $E(B-V)=0.51\pm 0.02$~mag
respectively from the blue edge of the RR Lyrae instability strip. For
comparison, the maps by Schlegel et al. (1998) give $E(B-V)=0.616$,
but these are known to overestimate reddening for objects close to the
galactic plane, like \object{NGC 6441}. Adopted $E(B-V)$\ values were
transformed into $E(V-K)$\ ones using the formula
$E(V-K)=2.75~E(B-V)$\ (Cardelli et al. 1989). Errors in these
temperatures are not easy to evaluate. Internal photometric errors in
the 2MASS $V-K$\ colors are generally small ($<0.04$~mag), leading to
random errors of $<25$~K. However, interstellar reddening is probably
variable in the field of \object{NGC 6441} (Layden et al. 1999; Pritzl et
al. 2001); four of the observed stars lie in the south-east sector of
the cluster, where Layden et al. found a reddening value in the range
$0.41<E(B-V)<0.54$. Assuming an $r.m.s.$ scatter of 0.05~mag in
$E(B-V)$\ (0.14~mag in $E(V-K)$), the random (star-to-star)
uncertainties in the effective temperatures may be as large as $\pm
80$~K. Systematic uncertainties are likely larger, including errors in
the Alonso et al. color-temperature calibration, in the photometric
calibration, and on the average reddenings we assumed. Hereinafter we
will assume possible systematic errors of $\pm 100$~K.

For stars lacking the 2MASS photometry, we used our $V-I$\
colors. Since these colors lack a proper photometric calibration,
we may use those stars having known $V-K$\ colors to construct a
correlation between the two colors, that turned out to be very narrow
(this is not unexpected, since both colors have a weak dependence on
metal abundance). The mean relation is $V-K=0.479+1.952~(V-I)$,
derived from more than 250 stars spanning over 4 magnitudes in $V-K$\
color; the $r.m.s.$ scatter around this mean relation is 0.098 mag
in ($V-K$). We could derive consistent temperatures also for
 stars with only $V-I$ colors, the errors due to uncertainties in the colors being
smaller than those related to the assumptions about reddening.

We may compare these temperatures derived from colors with those that
we could deduce from excitation equilibrium for Fe I lines. We found
that on average temperatures derived from Fe I excitation equilibrium
are lower by $70\pm 27$~K, with an $r.m.s.$ scatter for individual stars
of 95~K. This small difference can be attributed to several causes
(errors in the adopted temperature scale, inadequacies of the adopted
model atmospheres, etc.). On the whole we do not consider this
difference as important. The star-to-star scatter in the difference
between temperatures from colors and excitation equilibrium is not
much larger than the internal errors in the excitation temperatures
alone (68~K). This leaves space for errors in temperatures from
colors of about 66~K, to be attributed to the effect of differential
reddening. This corresponds to an r.m.s scatter of about 0.043~mag in
the reddening, compatible with results from photometry alone.
  
Surface gravities were obtained from the location of the stars in the
color-magnitude diagram. This procedure requires assumptions about
the distance modulus; (we adopted $(m-M)_V=16.33$\ from Harris (1996) for
the cluster members, while for the field bulge stars we assumed that
the stars are at the same distance of the galactic center: $7.9\pm
0.3$~kpc\ from Eisenhauer et al. (2003), the bolometric corrections
(from Alonso et al. 1999), and the masses (we assumed a mass of
0.9~M$_\odot$, close to the value given by isochrone
fittings). Uncertainties in these gravities are small for cluster
stars (we estimate internal star-to-star errors of about 0.06~dex, due
to the effects of possible variations in the interstellar absorption;
and systematic errors of about 0.15~dex, dominated by systematic
effects in the temperature scale).

We may compare these values for the surface gravities with those
deduced from the equilibrium of ionization of Fe.  On average,
abundances from Fe II lines are $0.01\pm 0.05$~dex smaller than that
derived from Fe I lines. The agreement is obviously very good. The
star-to-star scatter in the residuals is 0.17~dex, but it is as small
as 0.12~dex if only stars member of \object{NGC 6441} are considered; this
last value is close to the value expected considering that only three
to four Fe II lines were typically used in the analysis. The larger
spread obtained for field stars is not unexpected, since these stars
may be at different distances from the Sun.

Micro-turbulence velocities $v_t$\ were determined by eliminating
trends in the relation between expected line strength and abundances
(see Magain 1984). The error on the micro-turbulent velocities 
is determined by the change required on the derived $v_t$ 
to vary the expected line strength vs abundances slope by 1$\sigma$.  
This implies an expected random
error in the micro-turbulence velocities of $\pm 0.12$~km s$^{-1}$. 

Finally,the model atmospheres were iteratively chosen with  
model metal abundances in agreement with derived Fe
abundance. The adopted model atmosphere parameters are listed in
Table~\ref{t:uvesatmo}.

\begin{table}
\begin{center}
\caption{Iron abundances for stars observed with UVES}
\begin{tabular}{lcccccc}
\hline
Star  &    &   Fe I  &      &   &  Fe II  &     \\
      & n  &  [Fe/H] & $r.m.s.$  & n & [Fe/H]  & $r.m.s.$ \\
\hline
\multicolumn{7}{c}{Stars of \object{NGC 6441}}\\
\hline
7004050 & 52 & $-$0.45 & 0.18 & 4 & $-$0.49 & 0.18 \\
7004434 & 52 & $-$0.34 & 0.23 & 2 & $-$0.05 & 0.12 \\
7004463 & 57 & $-$0.38 & 0.18 & 4 & $-$0.56 & 0.46 \\
7004487 & 51 & $-$0.50 & 0.13 & 4 & $-$0.68 & 0.11 \\
8002961 & 59 & $-$0.28 & 0.16 & 4 & $-$0.34 & 0.16 \\ 
\hline
\multicolumn{7}{l}{$<{\rm [Fe/H] I}> = -0.39 \pm 0.04$, $r.m.s.$=0.09}\\
\multicolumn{7}{l}{$<{\rm [Fe/H]II}> = -0.42 \pm 0.11$, $r.m.s.$=0.24}\\
\hline
\multicolumn{7}{c}{Field stars}\\
\hline
6003734 & 63 & $-$0.25 & 0.20 & 6 & $-$0.16 & 0.17 \\ 
6004360 & 44 &   +0.27 & 0.24 & 7 &   +0.49 & 0.27 \\ 
6005308 & 63 & $-$0.17 & 0.21 & 8 &   +0.17 & 0.11 \\ 
6005341 & 60 &   +0.06 & 0.23 & 6 &   +0.01 & 0.28 \\ 
7004329 & 57 &   +0.12 & 0.16 & 7 & $-$0.08 & 0.24 \\ 
7004453 & 62 & $-$0.04 & 0.20 & 7 & $-$0.09 & 0.14 \\ 
8003092 & 46 & $-$0.25 & 0.23 & 3 & $-$0.44 & 0.23 \\
\hline
\end{tabular}
\label{t:uvesfe}
\end{center}
\end{table}

\begin{table*}
\begin{center}
\caption{Sensitivity of derived abundances to different sources of error.}
\begin{tabular}{lcccccccc}
\hline
Element   &Average &$T_{\rm eff}$&$\log{g}$& $[$A/H$]$ & $v_t$   &  EWs  &Total   &Total     \\
           &n. lines&         &         &         &         &       &Internal&Systematic\\
\hline
Variation  &        &    100  &    0.30 &   0.20  &   0.20  &       &       &       \\
Internal   &        &     50  &    0.06 &   0.07  &   0.12  & 0.194 &       &       \\
Systematic &        &    100  &    0.15 &   0.08  &   0.12  & 0.194 &       &       \\
\hline
$[$Fe/H$]${\sc i}   &  55.0  &$-$0.012 &  +0.063 &  +0.028 &$-$0.114 & 0.026 & 0.075 & 0.053 \\
$[$Fe/H$]${\sc ii}   &  ~2.7  &$-$0.210 &  +0.199 &  +0.046 &$-$0.046 & 0.119 & 0.167 & 0.262 \\
$[$O/Fe$]${\sc i}   &  ~1.8  &  +0.041 &  +0.057 &  +0.009 &  +0.101 & 0.145 & 0.161 & 0.155 \\
$[$Na/Fe$]${\sc i}  &  ~2.0  &  +0.106 &$-$0.069 &$-$0.025 &  +0.039 & 0.137 & 0.152 & 0.177 \\
$[$Mg/Fe$]${\sc i}  &  ~2.0  &$-$0.024 &$-$0.010 &$-$0.013 &  +0.084 & 0.137 & 0.149 & 0.141 \\
$[$Al/Fe$]${\sc i}  &  ~2.0  &  +0.087 &$-$0.060 &$-$0.028 &  +0.062 & 0.137 & 0.152 & 0.166 \\
$[$Si/Fe$]${\sc i}  &  ~2.2  &$-$0.102 &  +0.020 &$-$0.004 &  +0.079 & 0.131 & 0.151 & 0.168 \\
$[$Ca/Fe$]${\sc i}  &  11.4  &  +0.131 &$-$0.102 &$-$0.016 &$-$0.006 & 0.057 & 0.093 & 0.152 \\
$[$Sc/Fe$]${\sc ii}  &  ~2.8  &$-$0.017 &  +0.065 &  +0.008 &  +0.027 & 0.115 & 0.120 & 0.121 \\
$[$Ti/Fe$]${\sc i}  &  ~6.3  &  +0.167 &$-$0.050 &$-$0.014 &$-$0.061 & 0.078 & 0.123 & 0.187 \\
$[$Ti/Fe$]${\sc ii}  &  ~1.0  &$-$0.048 &  +0.071 &  +0.009 &  +0.095 & 0.194 & 0.206 & 0.205 \\
$[$V/Fe$]${\sc i}   &  12.8  &  +0.178 &$-$0.041 &$-$0.011 &$-$0.077 & 0.054 & 0.117 & 0.188 \\
$[$Cr/Fe$]${\sc i}  &  ~1.0  &  +0.150 &$-$0.025 &$-$0.015 &$-$0.055 & 0.194 & 0.212 & 0.246 \\
$[$Mn/Fe$]${\sc i}  &  ~3.0  &  +0.053 &$-$0.063 &  +0.007 &$-$0.026 & 0.112 & 0.120 & 0.128 \\
$[$Ni/Fe$]${\sc i}  &  23.5  &$-$0.028 &  +0.029 &$-$0.001 &  +0.032 & 0.040 & 0.054 & 0.052 \\
$[$Y/Fe$]${\sc i}   &  ~1.0  &  +0.231 &$-$0.044 &$-$0.018 &  +0.074 & 0.194 & 0.230 & 0.252 \\
$[$Zr/Fe$]${\sc i}  &  ~4.8  &  +0.207 &$-$0.025 &$-$0.014 &  +0.043 & 0.091 & 0.141 & 0.213 \\
$[$Ba/Fe$]${\sc ii}  &  ~3.0  &  +0.029 &  +0.029 &  +0.014 &$-$0.111 & 0.112 & 0.134 & 0.120 \\
$[$Eu/Fe$]${\sc ii}  &  ~1.8  &$-$0.004 &  +0.058 &  +0.009 &  +0.093 & 0.145 & 0.156 & 0.091 \\
\hline
\end{tabular}
\label{t:errorparam}
\end{center}
\end{table*}

\subsection{Fe Abundances}

Individual [Fe/H] values are listed in Table ~\ref{t:uvesfe}, as well
as averages over the whole sample. Reference solar abundances are as
in Gratton et al. (2003). Throughout our analysis, we use the same
line parameters discussed in Gratton et al. (2003); in particular,
collisional damping was considered using updated constants from Barklem
et al. (2000). We note that, for internal consistency, 
 we used four Fe{\sc II} lines for all member stars. In the case of star \# 7004463 
the lines yield considerably different Fe{\sc II} abundances, therefore the derived 
{\it r.m.s.} is larger than in other cases.

Table~\ref{t:errorparam} lists the impact of various uncertainties on
the derived abundances for the elements considered in our
analysis. Variations in parameters of the model atmospheres were
obtained by changing each of the parameters at a time for the star
\#7004487, assumed to be representative of all the stars
considered in this paper. The first three rows of the Table give the
variation of the parameter used to estimate sensitivities, the
internal (star-to-star) errors (for member stars of \object{NGC 6441}), and the
systematic errors (common to all stars) in each parameter. The first
column gives the average number of lines $n$\ used for each
element. Columns 3-6 give the sensitivities of the abundance ratios to
variations of each parameter. Column 7 gives the contribution to the
error given by uncertainties in EWs for individual lines: that is
$0.194/\sqrt{n}$, where 0.194 is the error in the abundance derived
from an individual line, as obtained by the median error over all the
stars (note that the errors in the EWs have much smaller impact for
the stars having better spectra, like \#7004487 itself). The two final
Columns give the total resulting internal and systematic errors, obtained by
summing quadratically the contribution of the individual sources of
errors, weighted according to the errors appropriate for each
parameter. For the systematic errors, the contribution due to
equivalent widths and to micro-turbulence velocities (quantities
derived from our own analysis), were divided by the square root of the
number of cluster members observed. Note that this error analysis does
not include the effects of covariances in the various error sources,
which are however expected to be quite small for the program stars.

Errors in Fe abundances from neutral lines are dominated by
uncertainties in the micro-turbulent velocity. We note that internal errors
in Fe {\sc I} abundances are dominated by errors in $v_t$.
We estimate total random
errors of 0.075 dex, and systematic errors of 0.053 dex. From
Table~\ref{t:uvesfe}, the average Fe abundance from all stars of
\object{NGC 6441} is [Fe/H]=$-0.39\pm 0.04$\ (error of the mean), with an
$r.m.s.$ scatter of 0.09\,dex from 5 stars. The first result of our
analysis is that the metallicity of \object{NGC 6441} is [Fe/H]=$-0.39\pm
0.04\pm 0.05$, where the first error bar includes the uncertainties 
related to star-to-star random errors, 
and the second one the systematic effects related 
to the various assumptions made in the analysis
 The offset between abundances given by neutral and
singly ionized Fe lines is only 0.03\,dex, supporting our analysis. The
star-to-star scatter (0.09\,dex) is slightly larger than expected on
the basis of our error analysis (0.075 dex); the scatter for Fe II is
larger, but in good agreement with expectations based on our error
analysis. While this difference is not significant, given the small
number of stars, it is possible that stars \#7004050 and \#7004487 are
indeed slightly more metal-poor (by about 0.15\,dex) than the
average of the remaining three cluster members. It is also worth
mentioning that star \#7004050 is also slightly bluer than the other
member stars of \object{NGC 6441}, by about 0.08 mag in the $V-I$\ color: this
is roughly the color difference expected for the metallicity
difference found in our analysis. More data on a wider sample of stars
of \object{NGC 6441} are clearly required to definitely set the issue about an
intrinsic spread in metal abundances in this cluster.

To our knowledge, this is the first determination of the metal
abundance of \object{NGC 6441} from high dispersion spectra of individual
stars. We found in the literature only three previous metallicity
determinations for \object{NGC 6441}. The first two are based on integrated
light. Zinn \& West (1984) used the Q39 index to obtain a metallicity
of [Fe/H]=$-0.59\pm 0.15$. A somewhat higher metallicity of
[Fe/H]=$-0.47\pm 0.12$\ dex was obtained by Armandroff \& Zinn (1988)
from spectroscopy of the Ca II IR triplet, calibrated on the
Zinn \& West scale. Armandroff and Zinn averaged these two values to
produce the value of [Fe/H]=$-0.53\pm 0.11$. That is listed by
Harris (1996) and usually adopted in the discussion of this
cluster. Very recently, Clementini et al. (2005) have measured metal
abundances for a few RR Lyrae in \object{NGC 6441} using a variant of the
$\Delta S$\ method, obtaining [Fe/H]=$-0.69\pm 0.06$\ on the Zinn \&
West scale, and [Fe/H]=$-0.41\pm 0.06$\ on the Carretta \& Gratton
(1997) scale, with a quite considerable star-to-star scatter of more
than 0.3 dex ($r.m.s.$). The present metal abundance should be compared
with that on the Carretta \& Gratton scale, which is based on an
analysis method quite consistent with that here considered. Indeed,
this last comparison is quite good. Finally, we note that if we add
the point of \object{NGC 6441} to the correlation between abundances from high
dispersion spectra by our group (see Carretta et al. 2001) and those
by Zinn \& West (1984), a linear relationship between the two scales
would be favored, rather than a cubic one as considered by Carretta
et al. (2001).

\begin{figure}
\includegraphics[width=8.8cm]{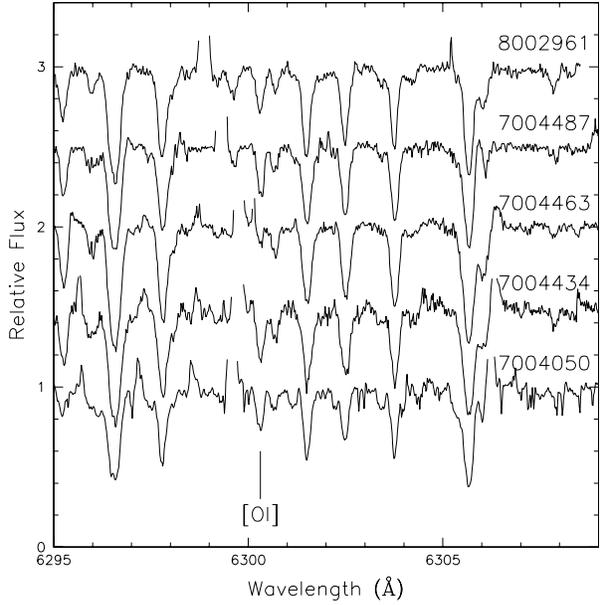}
\caption[]{Spectral region including the [OI] line at 6300.3~\AA\ 
for the member stars of \object{NGC 6441}. The spectra have been corrected for 
the individual radial velocities and offset vertically for clarity}
\label{f:ona1}
\end{figure}

\begin{figure}
\includegraphics[width=8.8cm]{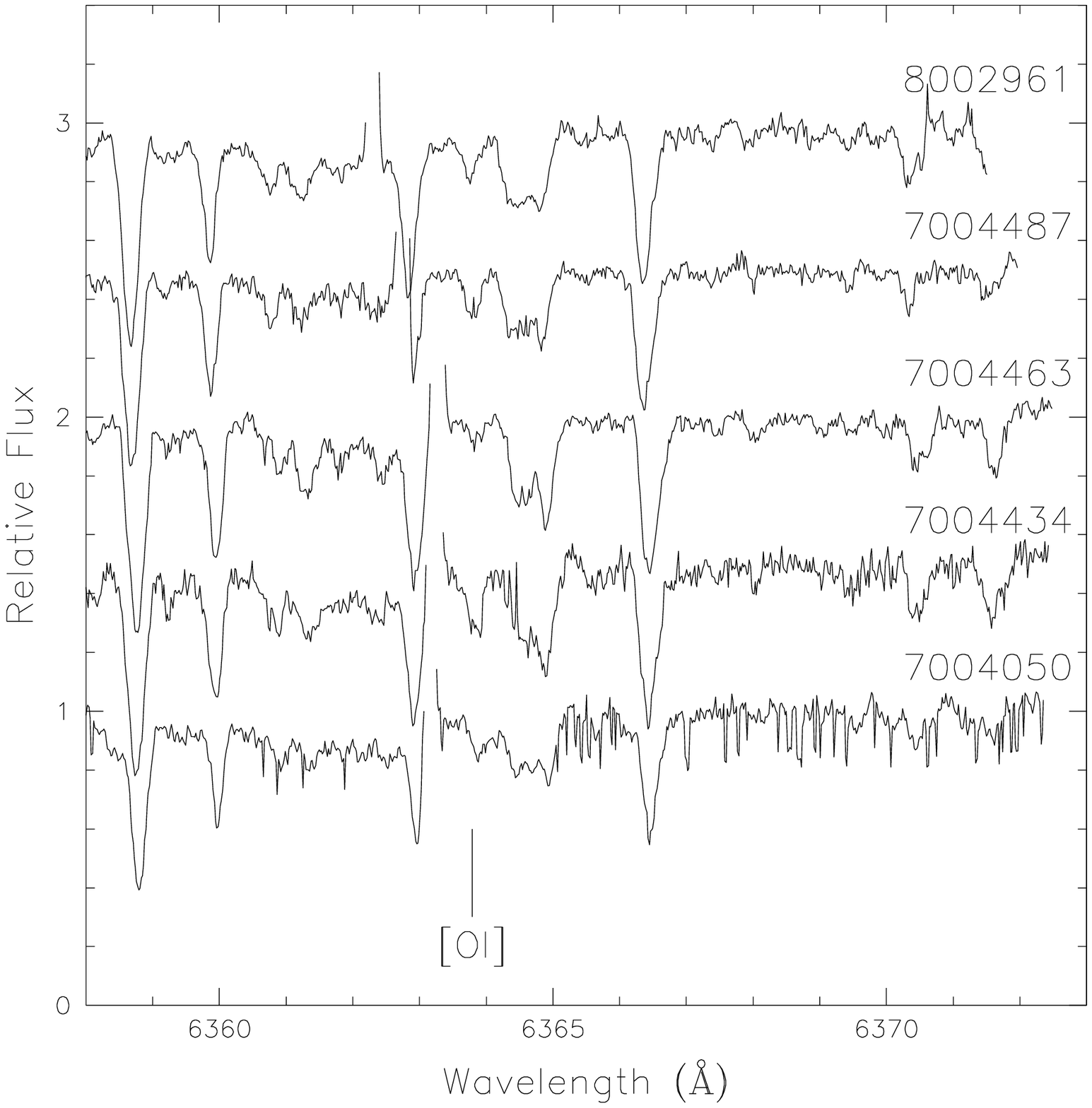}
\caption[]{Spectral region including the [OI] line at 6363.8~\AA\ for 
the member stars of \object{NGC 6441}. The spectra have been corrected for the 
individual radial velocities and offset vertically for clarity}
\label{f:ona2}
\end{figure}

\begin{figure}
\includegraphics[width=8.8cm]{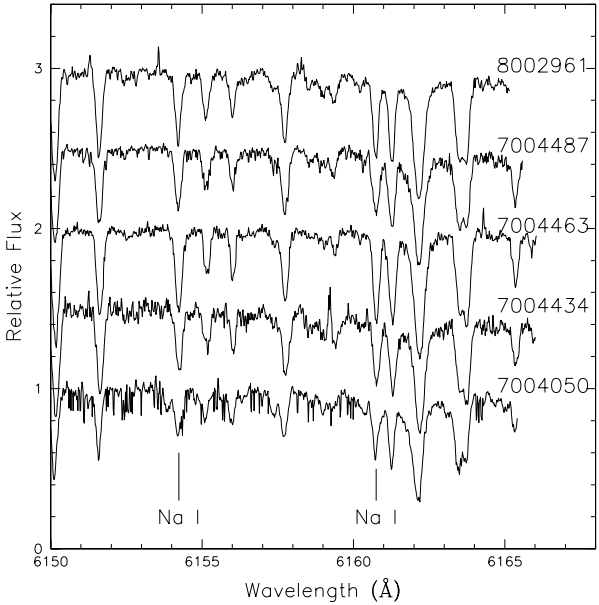}
\caption[]{Spectral region including the Na I doublet at 6154-60~\AA\ 
for the member stars of \object{NGC 6441}. The spectra have been corrected for 
the individual radial velocities and offset vertically for clarity}
\label{f:ona3}
\end{figure}

\begin{figure}
\includegraphics[width=8.8cm]{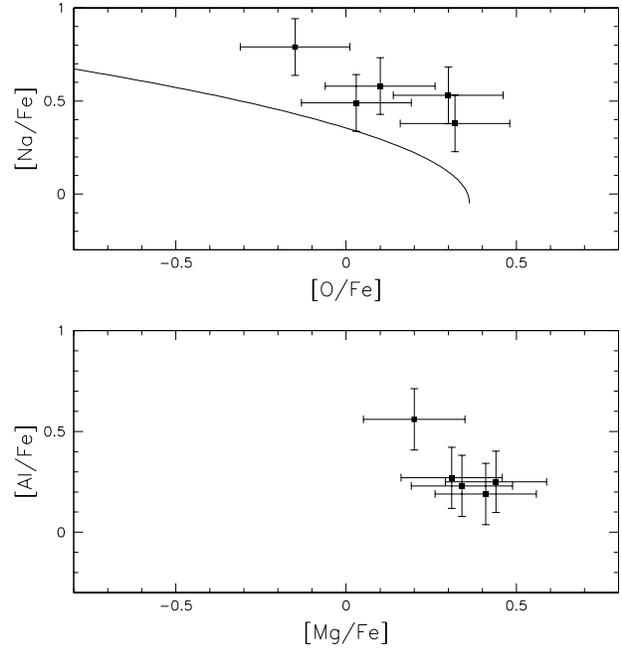}
\caption[]{Upper Panel: [Na/Fe] ratio as a function of [O/Fe], 
for stars member of \object{NGC 6441}, the curve indicates the mean 
[O/Fe] vs [Na/Fe] locus for a collection of $\sim$20 Globular clusters 
(Carretta et al. 2006). Lower panel:
[Al/Fe] ratio as a function of [Mg/Fe], for the same stars.}
\label{f:ona4}
\end{figure}

\begin{figure*}
\includegraphics[width=13cm]{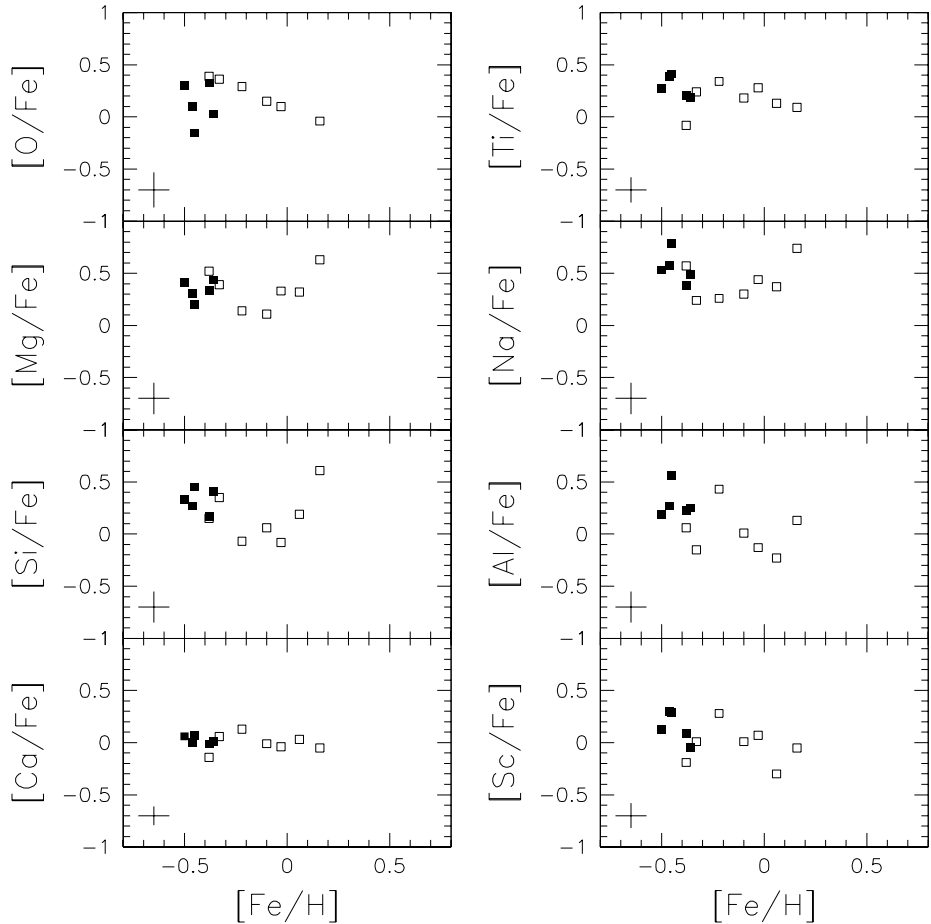}
\caption[]{Run of abundance ratios versus metallicity [Fe/H] for the 
analyzed stars. Filled symbols are stars member of \object{NGC 6441}; open 
symbols are field stars. Error bars are at bottom left of each panel. 
On the left, from top to bottom: [O/Fe], [Mg/Fe], [Si/Fe], and [Ca/Fe]. 
On the right, from top to bottom: [Ti/Fe] (from neutral lines alone), 
[Na/Fe], [Al/Fe], and [Sc/Fe]}
\label{f:6441a1}
\end{figure*}

\begin{figure*}
\includegraphics[width=13cm]{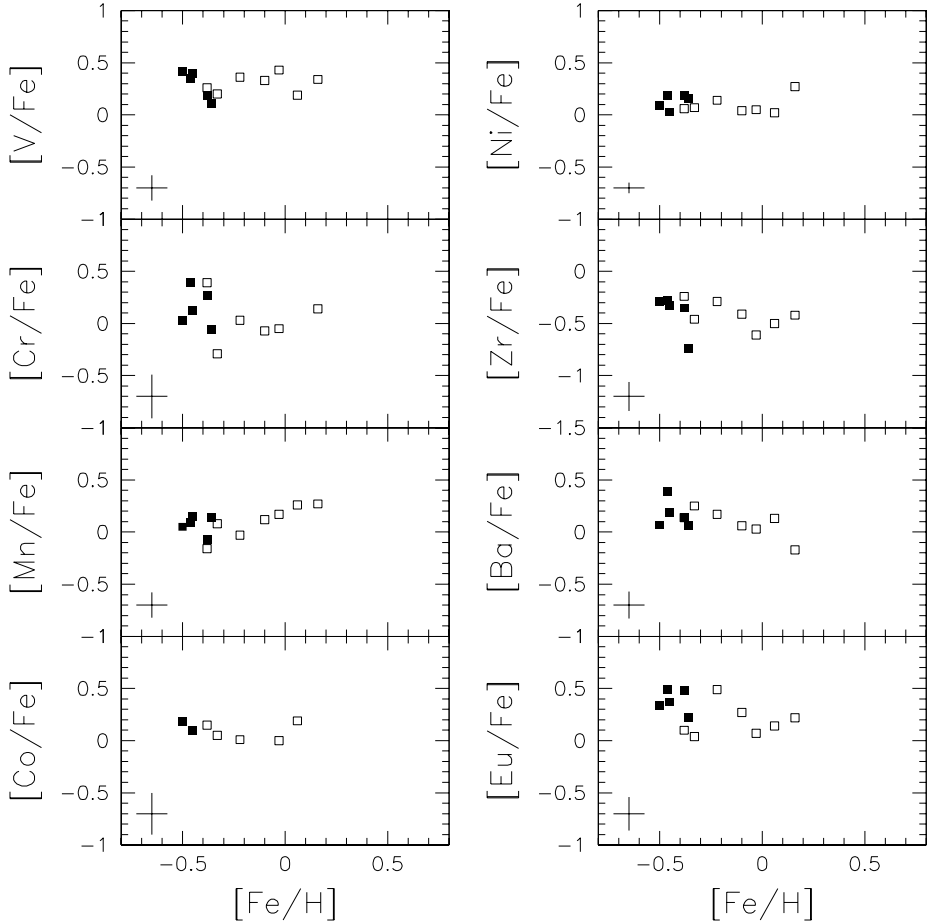}
\caption[]{Run of abundance ratios versus metallicity [Fe/H] for the 
analyzed stars. Filled symbols are stars member of \object{NGC 6441}; open 
symbols are field stars. Error bars are at bottom left of each panel. 
On the left, from top to bottom: [V/Fe], [Cr/Fe], [Mn/Fe], and [Co/Fe]. 
On the right, from top to bottom: [Ni/Fe], [Zr/Fe], [Ba/Fe], and [Eu/Fe]}
\label{f:6441a2}
\end{figure*}

\begin{table*}
\begin{center}
\caption{Abundances for \object{NGC 6441} members}
\scriptsize
\begin{tabular}{lccccccccccccccccc}
\hline
Element &\multicolumn{3}{c}{7004050}
        &\multicolumn{3}{c}{7004434}
        &\multicolumn{3}{c}{7004463}
        &\multicolumn{3}{c}{7004487}
        &\multicolumn{3}{c}{8002961}
        &Cluster&$r.m.s.$\\
&N&mean&$r.m.s.$&N&mean&$r.m.s.$&N&mean&$r.m.s.$&N&mean&$r.m.s.$&N&mean&$r.m.s.$&Average&\\
\hline
$[$Fe/H$]$   &53&$-$0.46&0.19&54&$-$0.38&0.23&59&$-$0.45&0.18&53&$-$0.50&0.13&59&$-$0.36&0.15&$-$0.43&0.06 \\
$[$O/Fe$]${\sc i}  &~2&  +0.10&0.08&~2&  +0.32&0.11&~1&$-$0.15&    &~2&  +0.30&0.14&~2&  +0.03&0.04&  +0.12&0.20 \\
$[$Na/Fe$]${\sc i} &~2&  +0.58&0.08&~2&  +0.38&0.10&~2&  +0.79&0.10&~2&  +0.53&0.00&~2&  +0.49&0.08&  +0.55&0.15 \\
$[$Mg/Fe$]${\sc i} &~2&  +0.31&0.01&~2&  +0.34&0.06&~2&  +0.20&0.02&~2&  +0.41&0.09&~2&  +0.44&0.08&  +0.34&0.09 \\
$[$Al/Fe$]${\sc i} &~2&  +0.27&0.08&~2&  +0.23&0.15&~2&  +0.56&0.23&~2&  +0.19&0.16&~2&  +0.25&0.21&  +0.30&0.15 \\
$[$Si/Fe$]${\sc i} &~1&  +0.27&    &~2&  +0.17&0.09&~2&  +0.45&0.17&~3&  +0.33&0.06&~2&  +0.41&0.16&  +0.33&0.11 \\
$[$Ca/Fe$]${\sc i} &12& ~~0.00&0.19&11&$-$0.01&0.20&10&  +0.07&0.15&12&  +0.06&0.22&12&  +0.01&0.16&  +0.03&0.04 \\
$[$Sc/Fe$]${\sc ii}&~3&  +0.30&0.24&~3&  +0.09&0.12&~3&  +0.29&0.20&~3&  +0.13&0.07&~3&$-$0.05&0.09&  +0.15&0.15 \\
$[$Ti/Fe$]${\sc i} &~5&  +0.39&0.09&~6&  +0.21&0.21&~6&  +0.41&0.18&~6&  +0.27&0.11&~7&  +0.19&0.15&  +0.29&0.10 \\
$[$Ti/Fe$]${\sc ii}&~1&  +0.50&    &~1&  +0.34&    &~1&  +0.18&    &~1&  +0.18&    &~1&  +0.43&    &  +0.33&0.14 \\
$[$V/Fe$]${\sc i}  &~9&  +0.35&0.17&15&  +0.18&0.18&14&  +0.40&0.17&12&  +0.42&0.15&11&  +0.11&0.14&  +0.29&0.14 \\ 
$[$Cr/Fe$]${\sc i} &~1&  +0.39&    &~1&  +0.27&    &~1&  +0.12&    &~1&  +0.03&    &~1&$-$0.06&    &  +0.15&0.18 \\
$[$Mn/Fe$]${\sc i} &~3&  +0.09&0.19&~3&$-$0.07&0.13&~3&  +0.15&0.20&~3&  +0.05&0.18&~3&  +0.14&0.12&  +0.07&0.09 \\
$[$Ni/Fe$]${\sc i} &21&  +0.18&0.16&23&  +0.18&0.17&23&  +0.03&0.20&25&  +0.09&0.20&24&  +0.16&0.14&  +0.13&0.07 \\
$[$Y/Fe$]${\sc i}  &~1&$-$0.34&    &~1&  +0.01&    &~1&  +0.31&    &~1&$-$0.11&    &~1&$-$0.10&    &$-$0.05&0.24 \\
$[$Zr/Fe$]${\sc i} &~4&$-$0.28&0.28&~5&$-$0.35&0.20&~5&$-$0.33&0.18&~5&$-$0.29&0.22&~5&$-$0.74&0.11&$-$0.40&0.19 \\
$[$Ba/Fe$]${\sc ii}&~3&  +0.39&0.21&~3&  +0.14&0.15&~3&  +0.19&0.18&~3&  +0.07&0.08&~3&  +0.06&0.18&  +0.17&0.13 \\
$[$Eu/Fe$]${\sc ii}&~1&  +0.49&    &~2&  +0.48&0.03&~2&  +0.37&0.07&~2&  +0.34&0.19&~2&  +0.22&0.04&  +0.32&0.11 \\
\hline
\end{tabular}
\normalsize
\label{t:abundgc6441}
\end{center}
\end{table*}

\begin{table*}
\begin{center}
\caption{Abundances for field stars}
\begin{tabular}{lcccccccccccc}
\hline
Element &\multicolumn{3}{c}{6003734}
        &\multicolumn{3}{c}{7004360}
        &\multicolumn{3}{c}{7005308}
        &\multicolumn{3}{c}{7005341}
        \\
&N&mean&$r.m.s.$&N&mean&$r.m.s.$&N&mean&$r.m.s.$&N&mean&$r.m.s.$\\
\hline
$[$Fe/H$]$   &62&$-$0.33&0.18&44&  +0.16&0.17&64&$-$0.22&0.21&62&$-$0.03&0.23\\
$[$O/Fe$]${\sc i}  &~2&  +0.36&0.01&~2&$-$0.04&0.15&~2&  +0.22&0.15&~1&  +0.10&    \\
$[$Na/Fe$]${\sc i} &~2&  +0.24&0.13&~2&  +0.74&0.19&~2&  +0.26&0.28&~2&  +0.44&0.10\\
$[$Mg/Fe$]${\sc i} &~2&  +0.39&0.13&~2&  +0.63&0.17&~2&  +0.14&0.34&~2&  +0.33&0.12\\
$[$Al/Fe$]${\sc i} &~2&$-$0.15&0.12&~2&  +0.13&0.17&~2&  +0.43&0.01&~2&$-$0.13&0.21\\
$[$Si/Fe$]${\sc i} &~2&  +0.35&0.16&~3&  +0.61&0.20&~2&$-$0.07&0.29&~2&$-$0.08&0.20\\
$[$Ca/Fe$]${\sc i} &13&$-$0.06&0.19&13&$-$0.05&0.17&11&  +0.13&0.17&11&$-$0.04&0.23\\
$[$Sc/Fe$]${\sc ii}&~2&  +0.01&0.12&~3&$-$0.05&0.32&~2&  +0.28&0.19&~3&  +0.07&0.46\\
$[$Ti/Fe$]${\sc i} &~7&  +0.24&0.20&~6&  +0.09&0.14&~7&  +0.34&0.17&~6&  +0.28&0.06\\
$[$Ti/Fe$]${\sc ii}&~1&$-$0.08&    &~1&  +0.39&    &~1&$-$0.02&    &~1&  +0.04&    \\
$[$V/Fe$]${\sc i}  &10&  +0.20&0.16&11&  +0.34&0.16&15&  +0.36&0.17&15&  +0.43&0.22\\ 
$[$Cr/Fe$]${\sc i} &~1&$-$0.29&    &~1&  +0.14&    &~1&  +0.03&    &~1&$-$0.05&    \\
$[$Mn/Fe$]${\sc i} &~3&  +0.08&0.13&~3&  +0.27&0.11&~3&$-$0.03&0.07&~3&  +0.17&0.11\\
$[$Co/Fe$]${\sc i} &~1&  +0.05&    &  &       &    &~1&  +0.01&    &~1& ~~0.00&    \\
$[$Ni/Fe$]${\sc i} &25&  +0.07&0.19&24&  +0.27&0.19&24&  +0.14&0.19&25&  +0.05&0.25\\
$[$Y/Fe$]${\sc i}  &~1&$-$0.23&    &~1&  +0.42&    &~1&$-$0.23&    &~1&$-$0.31&    \\
$[$Zr/Fe$]${\sc i} &~5&$-$0.46&0.20&~4&$-$0.42&0.24&~3&$-$0.29&0.09&~4&$-$0.61&0.17\\
$[$Ba/Fe$]${\sc ii}&~2&  +0.25&0.13&~3&$-$0.17&0.30&~3&  +0.17&0.13&~3&  +0.03&0.13\\
$[$Eu/Fe$]${\sc ii}&~1&  +0.04&    &~2&  +0.22&0.10&~2&  +0.49&0.09&~2&  +0.07&0.07\\
\hline
Element &\multicolumn{3}{c}{7004329}
        &\multicolumn{3}{c}{7004453}
        &\multicolumn{3}{c}{8003092}\\
&N&mean&$r.m.s.$&N&mean&$r.m.s.$&N&mean&$r.m.s.$\\
\hline
$[$Fe/H$]$   &58&  +0.06&0.16&64&$-$0.10&0.20&48&$-$0.38&0.23\\
$[$O/Fe$]${\sc i}  &  &       &    &~2&  +0.15&0.10&~2&  +0.39&0.07\\
$[$Na/Fe$]${\sc i} &~2&  +0.37&0.00&~2&  +0.30&0.15&~2&  +0.57&0.23\\
$[$Mg/Fe$]${\sc i} &~2&  +0.32&0.05&~2&  +0.11&0.02&~2&  +0.52&0.05\\
$[$Al/Fe$]${\sc i} &~2&$-$0.23&0.25&~2&$-$0.01&0.20&~2&  +0.06&0.17\\
$[$Si/Fe$]${\sc i} &~2&  +0.19&0.06&~2&  +0.06&0.10&~2&  +0.15&0.17\\
$[$Ca/Fe$]${\sc i} &11&$-$0.03&0.15&11&$-$0.01&0.17&10&$-$0.14&0.23\\
$[$Sc/Fe$]${\sc ii}&~3&$-$0.30&0.23&~3&  +0.01&0.20&~3&$-$0.19&0.24\\
$[$Ti/Fe$]${\sc i} &~6&$-$0.13&0.18&~7&$-$0.18&0.13&~5&$-$0.08&0.28\\
$[$Ti/Fe$]${\sc ii}&~1&$-$0.02&    &~1&  +0.33&    &~1&  +0.14&    \\
$[$V/Fe$]${\sc i}  &14&  +0.19&0.17&14&  +0.33&0.20&12&  +0.26&0.20\\ 
$[$Cr/Fe$]${\sc i} &  &       &    &~1&$-$0.07&    &~1&  +0.39&    \\
$[$Mn/Fe$]${\sc i} &~3&  +0.26&0.15&~3&  +0.12&0.16&~3&$-$0.16&0.14\\
$[$Co/Fe$]${\sc i} &~1&  +0.19&    &  &       &    &~1&  +0.15&    \\
$[$Ni/Fe$]${\sc i} &24&  +0.02&0.12&24&  +0.04&0.17&20&  +0.06&0.23\\
$[$Y/Fe$]${\sc i}  &~1&$-$0.31&    &~1&  +0.03&    &~1&$-$0.31&    \\
$[$Zr/Fe$]${\sc i} &~4&$-$0.50&0.20&~4&$-$0.41&0.14&~4&$-$0.24&0.22\\
$[$Ba/Fe$]${\sc ii}&~3&  +0.13&0.22&~3&  +0.06&0.18&~3&  +0.14&0.21\\
$[$Eu/Fe$]${\sc ii}&~2&  +0.14&0.22&~2&  +0.27&0.06&~2&  +0.10&0.07\\
\hline
\end{tabular}
\label{t:abundfield}
\end{center}
\end{table*}

\begin{table*}
\caption{Mean Abundances in Clusters and Group of Stars}
\begin{tabular}{lcccccccccccccc}
\hline
Element  &
\multicolumn{2}{c}{NGC 6441}&
\multicolumn{2}{c}{Metal-poor}&
\multicolumn{2}{c}{Metal-Rich}&
\multicolumn{2}{c}{NGC 6528}&
\multicolumn{2}{c}{NGC 6553}&
\multicolumn{2}{c}{Baade}&
\multicolumn{2}{c}{Terzan 7}\\
&\multicolumn{2}{c}{(1)}&
\multicolumn{2}{c}{(1)}&
\multicolumn{2}{c}{(1)}&
\multicolumn{2}{c}{(2)}&
\multicolumn{2}{c}{(3)}&
\multicolumn{2}{c}{Window (4)}&
\multicolumn{2}{c}{(5)}\\
&mean&$r.m.s.$&mean&$r.m.s.$&mean&$r.m.s.$&mean&$r.m.s.$&mean&$r.m.s.$&mean&$r.m.s.$&mean&$r.m.s.$\\
\hline
$[$Fe/H$]$   &$-$0.43&0.08&$-$0.31&0.08&$-$0.02&0.08&  +0.07&0.02&$-$0.16&0.08&$-$0.33&    &$-$0.59&0.07\\   
$[$O/Fe$]${\sc i}  &  +0.14&0.20&  +0.32&0.09&  +0.13&0.04&  +0.07&0.11&  +0.50&0.13&  +0.03&0.18&       &    \\
$[$Na/Fe$]${\sc i} &  +0.46&0.18&  +0.22&0.17&  +0.27&0.07&  +0.40&0.12&       &    &  +0.21&0.37&       &    \\
$[$Mg/Fe$]${\sc i} &  +0.34&0.09&  +0.35&0.19&  +0.25&0.12&  +0.14&0.09&  +0.41&0.10&  +0.35&0.14&$-$0.11&0.07\\
$[$Al/Fe$]${\sc i} &  +0.30&0.15&  +0.11&0.29&$-$0.12&0.12&       &    &       &    &       &    &       &    \\
$[$Si/Fe$]${\sc i} &  +0.33&0.11&  +0.14&0.21&  +0.06&0.14&  +0.36&0.07&  +0.14&0.18&  +0.18&0.24&  +0.07&0.09\\
$[$Ca/Fe$]${\sc i} &  +0.03&0.04&  +0.02&0.14&$-$0.01&0.04&  +0.23&0.06&  +0.26&0.09&  +0.14&0.17& ~~0.00&0.14\\
$[$Sc/Fe$]${\sc ii}&  +0.15&0.15&  +0.03&0.24&$-$0.07&0.20&$-$0.05&0.10&$-$0.12&0.18&  +0.29&0.20&       &    \\
$[$Ti/Fe$]${\sc i} &  +0.29&0.10&  +0.17&0.22&  +0.20&0.08&  +0.03&0.07&  +0.19&0.06&  +0.34&0.10&$-$0.05&0.07\\
$[$Ti/Fe$]${\sc ii}&  +0.33&0.14&  +0.01&0.11&  +0.12&0.19&       &    &       &    &       &    &       &    \\
$[$V/Fe$]${\sc i}  &  +0.29&0.14&  +0.27&0.08&  +0.32&0.12&$-$0.20&0.09&       &    &  +0.06&0.19&       &    \\
$[$Cr/Fe$]${\sc i} &  +0.15&0.18&  +0.04&0.34&$-$0.06&0.01& ~~0.00&0.04&  +0.04&0.09&$-$0.04&0.19&       &    \\
$[$Mn/Fe$]${\sc i} &  +0.07&0.09&$-$0.04&0.12&  +0.18&0.07&$-$0.37&0.07&       &    &       &    &       &    \\
$[$Co/Fe$]${\sc i} &  +0.14&0.07&  +0.07&0.07&  +0.10&0.13&       &    &       &    &       &    &       &    \\
$[$Ni/Fe$]${\sc i} &  +0.13&0.07&  +0.09&0.04&  +0.04&0.02&  +0.10&0.05&  +0.01&0.07&$-$0.04&0.08&$-$0.19&0.05\\
$[$Y/Fe$]${\sc i}  &$-$0.05&0.24&$-$0.26&0.05&$-$0.20&0.20&       &    &       &    &       &    &       &    \\
$[$Zr/Fe$]${\sc i} &$-$0.40&0.19&$-$0.33&0.12&$-$0.51&0.10&       &    &       &    &       &    &       &    \\
$[$Ba/Fe$]${\sc ii}&  +0.17&0.13&  +0.19&0.06&  +0.07&0.05&  +0.14&0.07&       &    &  +0.20&0.28&       &    \\
$[$Eu/Fe$]${\sc ii}&  +0.38&0.11&  +0.21&0.24&  +0.16&0.10&       &    &       &    &       &    &       &    \\
\hline
\end{tabular}
(1) This paper\\
(2) Carretta et al. (2001)\\
(3) Cohen et al. (1999)\\
(4) McWilliam \& Rich (1994)\\ 
(5) Sbordone et al. (2005) \\
\label{t:meanabund}
\end{table*}

\subsection{Abundances for other elements} 

Table~\ref{t:abundgc6441} lists the abundances for the individual
elements for stars member of \object{NGC 6441}, while Table~\ref{t:abundfield}
is for field stars. For each star and element, we give the
number of lines used in the analysis, the average abundance, and the
$r.m.s.$ scatter of individual values. The Na abundances include
corrections for departures from LTE, following the treatment by
Gratton et al. (1999). Abundances for the odd elements of the Fe group
(Sc, V, and Mn) were derived with consideration for the not negligible
hyperfine structure of these lines (see Gratton et al. 2003 for more
details). Finally, we note that telluric absorption lines were removed
from the spectra in the region around the [OI] lines. No attempt was
made to remove the strong auroral emission line; however, due to the
combination of the Earth and stellar motions at the epoch of
observations, the auroral emission line typically is 
about 0.7~\AA\ blueward to the stellar line, so that it does not
create problems at the resolution of the UVES spectra. The O
abundances were derived from equivalent widths: however, they were
later confirmed by spectral synthesis. We did not apply any correction
for the blending with the Ni~I line at 6300.339~\AA, nor for
formation of CO. The blending Ni~I line is expected to contribute
about 4~m\AA\ to the EW of the [OI] line, using the line parameters by
Allende Prieto et al. (2001); this corresponds to correcting the O
abundances about 0.05 dex downward. CO coupling should be strong at
the low effective temperature of the program stars. Unluckily, the
abundance of C is not determined. However, we expect that C is
strongly depleted in stars near the tip of the red giant branch, with
expected values of [C/Fe]$\sim -0.6$\ (Gratton et al. 2000). If the original 
C abundance in unevolved stars follows the Fe one (as usually observed in metal-rich
environments: see e.g. Gratton et al. 2000), then we expect typical
values of $[$C/O$]\sim -0.8$\ for stars in \object{NGC 6441}. Neglecting CO
formation, we should have underestimated the O abundances from
forbidden lines by $\sim 0.05$~dex. These two corrections should 
roughly compensate, however, they both are within the error bars of
the present determinations.

The last two columns of Table~\ref{t:abundgc6441} give the average abundance for the cluster, as
well as the $r.m.s.$ scatter of individual values around this mean
value. In general, the values for the scatter agree fairly well with
those estimated in our error analysis. Large scatters are however
determined for the light elements O, Na, Mg, and Al; the
elements participating in the so-called Na-O anticorrelation.
Inspection of Table~\ref{t:abundgc6441} reveals that while four stars
share a similar composition (high O and Mg abundances, low Na and Al
ones), one star (\#7004463) has much lower O and Mg abundances, and
higher Na and Al ones. Figures~\ref{f:ona1}, \ref{f:ona2}, and
\ref{f:ona3} show the spectral region around the two [OI] lines and
the Na doublet at 6154-60~\AA\ in the spectra of the stars of
NGC 6441. The weakness of the [OI] lines and the strength of the Na
doublet in the spectrum of star \#7004463 is quite obvious (note that
the five stars have all similar temperatures and reddenings). This
star seems a typical representative of the O-poor, Na-rich stars often
found among Globular cluster stars. Figure~\ref{f:ona4} shows the O-Na
and Mg-Al anticorrelations from our data set. The curve plotted in the upper panel 
is from Fig. 5 in Carretta et al. (2006) and indicates the typical behavior of 
[O/Fe] with respect to [Na/Fe] as defined by a collection of stars in 
about 20  Globular clusters. More extensive data
about the O-Na anticorrelation among stars in \object{NGC 6441} will be
discussed separately, on the basis of the GIRAFFE spectra taken
simultaneously with the UVES ones.

\subsection{Comparison between \object{NGC 6441} and the field stars} 

The runs of the abundance ratios with respect to Fe are plotted
against metal abundances in Figures~\ref{f:6441a1} and
\ref{f:6441a2}. In these plots, filled symbols are for member stars
of \object{NGC 6441}, while open symbols are for field stars. These figures
highlight that the field stars are a quite heterogeneous group. Three stars
(\#6003734, 7005308, and 8003092) are moderately metal-poor, with a
metal abundance only slightly larger than that of \object{NGC 6441}. They have a
composition quite similar to that of the cluster, though the last one
has a smaller excess of Si. The two first stars have rather large
absolute values of the radial velocities: this suggests that they are
old, the excess of Oxygen and $\alpha-$elements being possibly due to
a reduced contribution by type-Ia SNe. Three other stars (\#7005341,
7004329, and 7004453) have almost solar abundances. They have a low
absolute value of the radial velocity.
 Finally, star \#7004360 has a peculiar, high
metal content. Our analysis yielded a significant excess of the
$\alpha-$elements Mg and Si, but not of O, Ca, and Ti; also Na is
overabundant, but not Al. However, given the difficulties related to
the analysis of this spectrum, extremely line rich, we prefer not to
comment any more on this star.

Comparison between field (likely bulge) and cluster stars may be interesting and
meaningful, because most of the analysis concerns are reduced. In
particular, deficiencies of model atmospheres or departures from LTE
are expected to act similarly in stars having similar atmospheres. We
may for instance compare the stars of \object{NGC 6441} with the group of
the field "metal-poor" stars, that have similar overall metal
abundances (see Table~\ref{t:meanabund}). The two sets of stars
display a quite similar pattern of abundances, both characterized by
an excess of $\alpha-$elements; also the apparent deficiency of Ca, Y,
and Zr is similar. On this respect we notice that the three field stars 
(likely bulge) display similar Ca abundances ([Ca/Fe]$\sim -0.2$), and
similar abundances for Y ([Y/Fe]$\sim -0.5$) and Zr ([Zr/Fe]$\sim
-0.8$), suggesting that the low Ca, Y and Zr abundances are artifacts
of the analysis, rather than real features of these stars. We notice
that the present analysis makes the same assumptions of Gratton et
al. (2003), where we did not find a similar low Ca abundance in
(local) metal-rich dwarfs. It seems more likely that the problem is
limited to the analysis of bright giants. All Ca lines used here are
strong, with $EWs\geq 100$~m\AA. They form at tiny optical depths, and
are potentially sensitive to details of the model atmospheres as well
as to departures from LTE. However, the differential comparison with
the non metal rich bulge stars suggests that \object{NGC 6441} and the field metal-poor stars
considered in this analysis have a small but clear excess of Ca, and
roughly normal abundances of Y and Zr.

The main difference between the stars of \object{NGC 6441} and the field
moderately metal-poor stars, concerns the elements involved in the
p-capture process of O, Na, Mg, and Al. Even leaving aside the obvious
case of the O-poor, Na-rich star \#7004463, the stars of \object{NGC 6441} are
systematically more O and Mg poor, and Na and Al rich than the similar
field stars. This suggests that even the more normal star might have
been somewhat polluted by material processed through H-burning at high
temperature. This is well confirmed by a comparison of the [O/Na]
abundance ratios in \object{NGC 6441} with those typical of other Globular
cluster stars (see Fig. 5 in Carretta et al. 2006). The other analyzed elements
appear much more similar between \object{NGC 6441} and the moderately metal-poor
field stars, although the abundance ratios to Fe are generally higher
in the stars of \object{NGC 6441} by about 0.05-0.1 dex, which might be
explained collectively by assuming some difference in Fe. This might
suggest the existence of some subtle difference in the
nucleosynthesis, although this result is at the limit of the
observational and analysis uncertainty, and we do not give
much weight to it.

Among the n-capture elements, stars in \object{NGC 6441} have [Ba/Fe] abundance
ratios similar to that of field stars; the [Eu/Fe] abundance ratios are
on average larger than the putative metal rich bulge stars (by about 0.2 dex). The
larger [Eu/Ba] ratios may be again explained as due to larger
contributions by core collapse SNe.

\subsection{Comparison with other bulge objects}

Comparisons with abundances from the literature must be considered
with more caution, due to possible systematic differences in the analysis. In
Table~\ref{t:meanabund} we give average abundances for two other
bulge clusters analyzed with a similar method 
(NGC 6528: Carretta et al. 2001; \object{NGC 6553}: Cohen et
al. 1999), Baade's Window (McWilliam \& Rich 1994), and for the
cluster Terzan 7, likely belonging to the Sagittarius Dwarf Galaxy
(Sbordone et al. 2005). The pattern of abundances seen in the various
bulge clusters is quite similar. The main differences concern: (i) the
light elements involved in the p-capture processes (O, Na, and
Mg). However, they may result from small number statistics, given the
large star-to-star spread observed within each cluster. (ii) Ca, for
which we think our abundance is not representative of the cluster real
abundance (the stars discussed here are much cooler than those
analyzed in \object{NGC 6528} and
NGC 6553). (iii) Mn, for which we derive a nearly solar ratio to Fe in
NGC 6441, while Carretta et al. derived a low abundance in
NGC 6528. This difference in Mn abundance might indicate a different
chemical history. Mn is known to be under-abundant in quite metal-rich
stars of the Sagittarius dwarf spheroidal (McWilliam et al. 2003);
however the composition of \object{NGC 6528} (and \object{NGC 6441}) is also clearly
different from that of Terzan 7 and other stars in the Sagittarius
Dwarf Galaxy, since in these last cases the $\alpha-$elements are not
at all overabundant. All these facts suggest that various bulge
clusters had different chemical histories, possibly being originated
in different fragments of our Galaxy.

\section{Conclusion}
This paper presents the first high resolution spectroscopic analysis 
of individual \object{NGC 6441} stars.
A total of thirteen RGB stars were observed using FLAMES/UVES, only 
five of them, on the basis of their measured 
radial velocities, positions and chemical compositions are very likely 
 members of \object {NGC 6441}.
The iron abundance measured for the cluster stars is [Fe/H]=$-0.39\pm0.04\pm0.05$\,dex,
 somewhat higher than those reported in the literature (see e.g. 
Zinn \& West 1984 and Armandroff \& Zinn, 1988) which are however based on 
lower resolution data. 
In a more recent paper, Clementini et al. (2005) measured the metal abundance 
of  \object {NGC 6441} RR Lyrae stars finding an average metallicity of 
[Fe/H]=$-0.41\pm0.06$\,dex which is perfectly consistent 
 with our results. 

Some of the results obtained seem to hint that \object{NGC 6441} was characterized by 
slightly different nucleosynthetic processes. In fact, 
in the cluster stars O and Mg are systematically lower and Na and Al higher
than in the field stars, and the [O/Na] ratios measured in \object{NGC 6441} 
are different from those typical of other clusters. 

We find a very small metallicity spread among the stars in our 
sample which belong to the cluster, only marginally larger than the level 
expected on the basis of the observational error, suggesting an homogeneous composition.
However, our sample is far too small to be representative of the cluster distribution
and, on the other hand, Clementini et al. (2005), find a scatter as high as 0.3\,dex. 
The derived homogeneity could be just due to limited sampling and thus    
no definite conclusion can be drawn on the basis of the present data. 
The GIRAFFE/FLAMES data presented by Gratton et al. (2006 in preparation) 
as well as the near-IR NIRSPEC ones in Origlia et al. (2006 in preparation) will 
hopefully address this issue.

\begin{acknowledgements}
We wish to thank our anonymous referee for his/her comments and suggestions.
This research has been funded by PRIN 2003029437 "Continuit\`a e
discontinuit\`a nella formazione della nostra Galassia" by the Italian
MIUR. This publication makes use of data products from the Two Micron
All Sky Survey, which is a joint project of the University of
Massachusetts and the Infrared Processing and Analysis
Center/California Institute of Technology, funded by the National
Aeronautics and Space Administration and the National Science
Foundation
\end{acknowledgements}

\end{document}